\documentclass[jcp,aps,showpacs,twocolumn,supergroupedaddress,epsfig,amsmath,amssymb]{revtex4}
\usepackage{epsfig,amsmath,amsthm,amssymb,bm,epsf,graphics,psfrag,verbatim,subfigure,framed}
\usepackage[makeroom]{cancel}
\usepackage{mathtools}
\usepackage[usenames,dvipsnames]{color}
\usepackage{empheq}
\usepackage{bbm}
\usepackage{mathrsfs}
\usepackage{amsfonts}
\usepackage{color} 
\usepackage{pbox}
\usepackage{xcolor} 
\usepackage{multirow}
\def\av{\mathbf{a}}

\newcommand{\be}{\begin{equation}}
\newcommand{\ee}{\end{equation}}
\newcommand{\ba}{\begin{eqnarray}}
\newcommand{\ea}{\end{eqnarray}}
\newcommand{\bw}{\begin{widetext}}
\newcommand{\ew}{\end{widetext}}

\newcommand{\Pv}{{\bf{P}}}

\newcommand{\rv}{{\mathbf{r}}}

\newcommand{\kv}{{\bf k}}
\newcommand{\pv}{{\bf p}}

\newcommand{\BSL}[1]{\textcolor{blue}{[BSL: #1]}}

\begin{document}

\title{Quenched fluctuation-induced force arising from polarization disorder}
\author{Bing-Sui Lu}
\email{binglu@aus.edu}
\affiliation{College of Arts and Sciences, American University of Sharjah, P.O. Box 26666, Sharjah, United Arab Emirates}
\date{\today}

\begin{abstract}
We investigate the zero-temperature behavior of the fluctuation force induced by the quenched disorder of electric dipoles frozen randomly into a material. 
Examples of such materials include relaxor ferroelectrics. 
In terms of the setup and geometry, we focus on a layered system comprising two coplanar semi-infinite slabs separated by a distance $\ell$ as well as a system comprising a neutral atom in the vacuum located at a distance $\ell$ above the surface of a semi-infinite slab. For both systems, we consider the cases where the quenched random dipolar disorder occurs inside the bulk as well as on the surface of the slabs. In all of these cases, we find that the bulk (surface) dipolar disorder-induced force grows with the mean square quenched electric dipole moment per unit volume (area). The bulk (surface) dipolar disorder-induced pressure between two semi-infinite single-layered slabs decays with $\ell^{-3}$ ($\ell^{-4}$), whereas the bulk (surface) dipolar disorder-induced force on an atom in the vacuum near a slab containing the disorder decays with $\ell^{-4}$ ($\ell^{-5}$). 
We also find that the quenched dipolar disorder-induced force between two coplanar slabs can be repulsive in a three-layered dielectric system that obeys a zero-frequency analogue of the Dzyaloshinskii-Lifshitz-Pitaevskii condition, and serve to enhance the ``nanolevitation effect" if the latter is present in the disorder-free system.
\end{abstract}

\maketitle

\section{Introduction}

Unlike the situation with bodies separated by macroscopic distances, material surfaces separated by distances ranging from the order of nanometers to the order of microns can experience a varied host of interaction forces that are induced by fluctuations, with the fluctuations being of different origins. For example, the thermal fluctuations of the ions in the bathing solution in which colloids are immersed can give rise to electrostatic correlation forces between them~\cite{podgornik1989,kanduc2010,naji2013}. Surfaces can also interact via charge regulation forces if they possess chemical groups that can gain or lose protons thereby causing their surface charge densities to fluctuate in response to the changing local  environment as the surfaces approach each other~\cite{podgornik2018,avni2020,bozic2018}. 
Casimir and van der Waals (vdW) forces can also become dominant between electrically neutral materials separated by vacuum gaps of nanoscale width and for neutral atoms positioned in the vacuum above electrically neutral materials~\cite{bordag2009,lifshitz1955,casimir1948,wylie1985,bloch2005,laliotis2021,lujcp2015,lujcp2016,deanjcp2012,rezvani2012,podgornikjcp2006}. Being attractive, these forces can give rise to stiction~\cite{maboudian1997}. On the other hand, owing to the fabrication process and/or adsorption by contaminant particles, a material can be overall electrically neutral, but contain random parasitic monopolar charges that are frozen into and distributed across the surface or inside the bulk. Such quenched monopolar charge disorder can further give rise to a  force between material surfaces which is attractive, albeit one that decays more weakly compared to Casimir/vdW forces and can in fact compete with the latter if the separation between the surfaces is large enough~\cite{camp1991,speake1996,ke2023,speake2003,behunin2012,fosco2013,behunin2014b,naji2005,podgornik2006,dean2011,dean2012,rezvani2012,sarabadani2010,naji2010,lu2025}. 

\    

Even if parasitic monopolar charge disorder can be neglected, polar molecules can adsorb onto material surfaces via physisorption or chemisorption, resulting in electric dipoles randomly distributed across the surface~\cite{tan2020,jaroenapibal2016,zhao2005,nakao2015}. There are also materials such as relaxor ferroelectrics and dipole glasses which intrinsically contain polar nanoregions which are locally neutral but behave effectively as electric dipoles, being generated by substitutional disorder, with the polar nanoregions being randomly distributed in the bulk of the material~\cite{kanzig1964,burns1983a,vugmeister1990,bokov2006,shvartsman2008,shvartsman2013,yokota2007,cai2015}. 
At temperatures well below the Vogel-Fulcher temperature (i.e., the temperature at which it takes an infinitely long time for the dipoles to relax)~\cite{bokov2006}, 
the random electric dipoles in a relaxor ferroelectric such as KLT (or potassium lithium tantalate, $\rm{K_{1-x}Li_xTaO_3}$)~\cite{cai2015} can be frozen into a glassy disordered phase, giving rise to a zero mean polarization but a nonzero mean square polarization. As we will show in this paper, the presence of the quenched random polarizations can also give rise to a disorder-induced fluctuation force. Depending on the relative dielectric permittivities of the dielectric materials and their intervening medium, this disorder-induced  force between two coplanar dielectric slabs can be attractive or repulsive. 
As was noted in the literature on the force behavior induced by quenched {\em monopolar} charge disorder, the disorder-induced  force arises essentially from the interactions of random monopolar charges with their image charges, since the interactions between different monopolar charges vanish when averaged over the random statistics of the charges~\cite{dean2012,lu2025}. In a similar vein, we may regard the disorder-induced  force arising from quenched dipolar charge disorder as essentially being the result of random dipoles interacting with their own image dipoles. 
The investigation of forces induced by quenched dipolar charge disorder can be relevant to the design of microactuators, microrobots and flexoelectric microdevices that involve dipole glasses and relaxor ferroelectrics~\cite{duval2007,lin2026,ivan2011,yi2025}. Stiction is a perennial issue, and results from such an investigation may be useful in designing devices that reduce stiction.  

\

In this paper, we investigate the behavior of the disorder-induced force arising from quenched random electric dipoles. For our purpose, we consider a layered system comprising two coplanar slabs separated by a distance $\ell$ as well as a system comprising a neutral atom in the vacuum located at a distance $\ell$ above the surface of a slab. For both systems, we investigate the cases where the quenched random dipolar disorder occurs inside the bulk as well as on the surface of the slabs, and obtain the corresponding behavior of the disorder-induced forces. 
In particular, we consider a system comprising two slabs separated by an intervening medium, one of the slabs being single-layered whereas the other is a semi-infinite substrate coated by a film containing quenched random electric dipoles. We investigate how the quenched bulk dipolar disorder-induced force interpolates between two asymptotic scaling behaviors as the thickness of the film is varied from being much smaller than the width of the intervening gap to being much larger. 
The dipolar disorder-induced force acting on an atom is obtained by the rarefaction method~\cite{bordag2009,lifshitz1955,lu2025b}, i.e., replacing the dielectric permittivity by that of a homogeneous, non-interacting gas of atoms. 
We will consider glassy disordered systems at zero temperature, for which thermal dipolar fluctuations can be neglected.  

\

The present paper is organized in the following manner. We start in Section II with the electrostatic energy of the quenched random polarizations and the relevant electrostatic Green functions. In Section III, we derive expressions for the disorder-induced pressure between dielectric slabs and the disorder-induced  force acting on an atom above a dielectric slab that are induced by the presence of quenched random electric dipoles in the bulk of a given layer, whilst in Section IV, we derive analogous expressions for the case where the quenched random electric dipoles reside on the surface of a layer that is adjacent to the intervening gap. We present and discuss the results of our investigations in Section V, and we provide a summary and conclusion in Section VI.

\section{Model and Green functions}

\begin{figure}[h]
    \centering
      \includegraphics[width=0.48\textwidth]{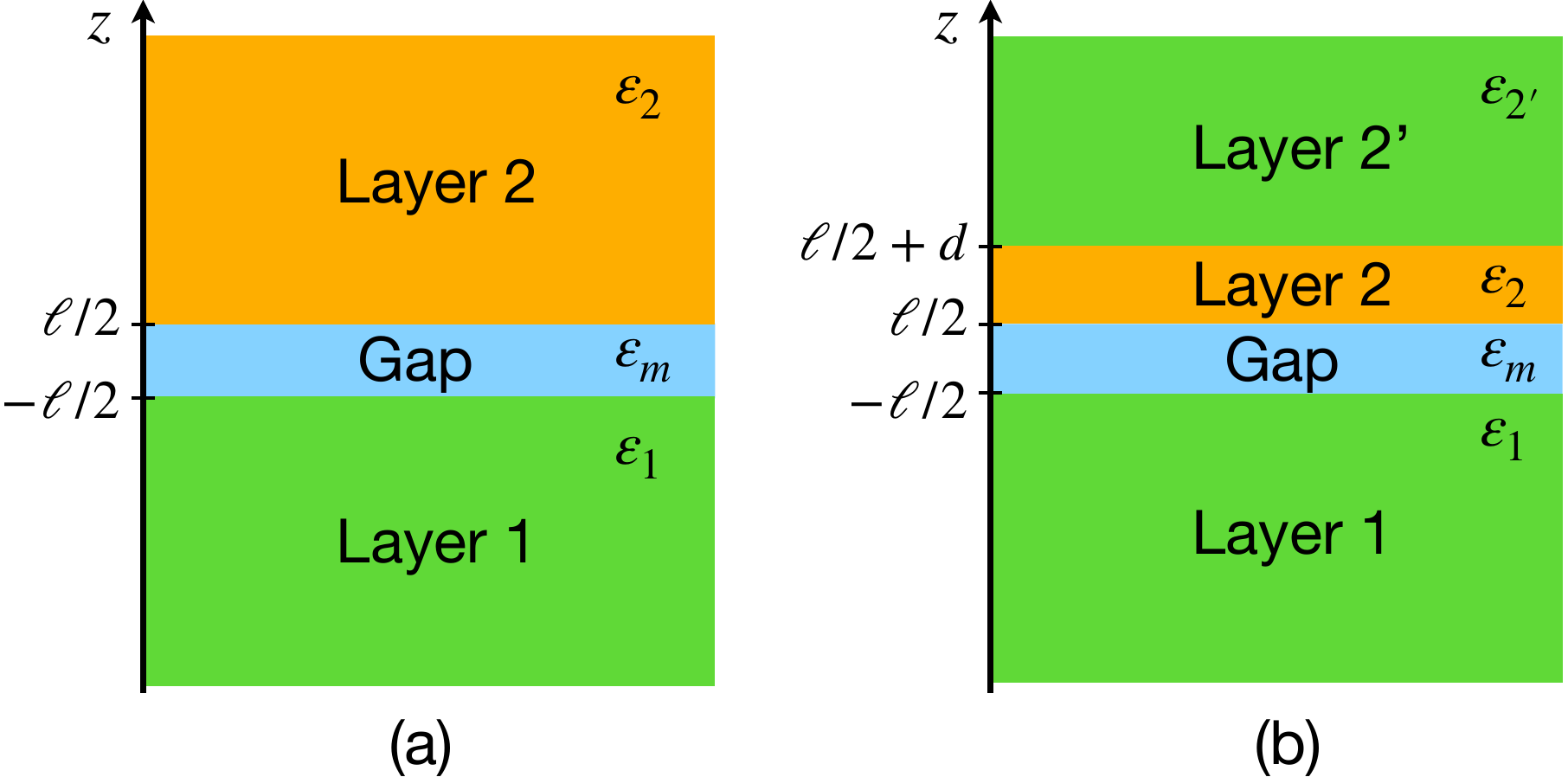}
      \caption{(a)~A pair of coplanar semi-infinite single-layered slabs with dielectric permittivities $\varepsilon_1$ and $\varepsilon_2$ separated by an intervening gap occupied by a medium of dielectric permittivity $\varepsilon_m$. Layer 2 has quenched dipolar disorder, whilst layer 1 is disorder-free. (b)~A system consisting of a single-layered slab and a bilayered slab: layer 1 is semi-infinite with dielectric permittivity $\varepsilon_{1}$. Layer 2 has a thickness $d$ and dielectric permittivity $\varepsilon_2$, whilst layer $2'$ is semi-infinite and has a dielectric permittivity $\varepsilon_{2'}$. Layer 2 contains quenched dipolar disorder, whilst layers 1 and $2'$ are disorder-free.} 
      \label{setup-slab}
\end{figure}

To investigate the behavior of the disorder-induced  pressure between two coplanar slabs that arises from the presence of quenched random polarizations, we will focus on two types of systems: (i)~one which involves a pair of semi-infinitely thick single-layered slabs separated  by a gap of width $\ell$ (as depicted in Fig.~\ref{setup-slab}(a)), and (ii)~another which involves a single-layered slab that is coplanar with a bilayered slab and also separated by a gap of width $\ell$ (which is depicted in Fig.~\ref{setup-slab}(b)). 
We will assume that the layers are dielectrically homogeneous (i.e., with scalar dielectric permittivities) and there is no mobile charge screening in any of the layers. 
The layers are labeled by the index $\lambda$. 
For instance, if we are considering a system such as that described by Fig.~\ref{setup-slab}(a), ${\lambda } = 1$ would refer to the bottom (green-colored) layer, ${\lambda } = 2$ to the top (ochre-colored) layer, and ${\lambda } = m$ to the medium occupying the intervening gap (colored light blue in Fig.~\ref{setup-slab}). 
By defining the polarization density $\Pv_{\lambda }(\rv)$ via 
\be
\Pv_{\lambda }(\rv) \equiv \sum_I \pv_{I{\lambda }} \delta(\rv - \rv_{I{\lambda }}), 
\label{P-define}
\ee
where $\pv_{I{\lambda }}$ is the $I$th electric dipole in layer ${\lambda }$, and introducing the Green function $G(\rv,\rv')$, which obeys 
\be
-{\bm \nabla} \!\cdot\! [ \varepsilon(\rv) {\bm \nabla} G(\rv, \rv') ] = 4\pi \delta(\rv-\rv'),  
\label{green-eq1}
\ee
where $\varepsilon(\rv)$ is the local dielectric permittivity, 
we can express the electrostatic energy of the quenched random polarizations by 
\be
U_E = 
\frac{1}{2} \sum_{{\lambda },{\lambda' }} \int \! d^3\rv \! \int \! d^3 \rv' 
P_{{\lambda } i}(\rv) P_{{\lambda' } j}(\rv') \nabla_{i} \nabla'_{j} G(\rv,\rv'). 
\label{UE-dipole}
\ee
Here, $i,j = x,y,z$ are Cartesian indices. The above form of the energy is valid at slab-slab (or atom-slab) separations that are much larger than the typical size of the quenched random dipoles. For a relaxor ferroelectric, this would be the typical size of a polar nanoregion, which can be as small as 1.5 nm and as large as 18 nm~\cite{bokov2006}.

As the translation symmetry is broken along the $z$ direction (i.e., the direction perpendicular to the slab surface), we can express the Green function $G(\rv,\rv')$ in terms of a two-dimensional Fourier transform, viz., 
\be
G(\rv,\rv') = \int \!\! \frac{d^2 \kv_\parallel}{(2\pi)^2} \, 
e^{i\kv_\parallel \cdot (\rv_\parallel - \rv'_\parallel)} 
\widetilde{G}(\kv_\parallel, z,z'), 
\label{G-fourier}
\ee
where $\kv_\parallel = (k_x, k_y)$ is the two-dimensional wavevector parallel to the plane of the slab surface. 
By substituting Eq.~(\ref{G-fourier}) into Eq.~(\ref{green-eq1}), we obtain 
\be
\left[ -\frac{\partial}{\partial z} \left( \varepsilon(z) \frac{\partial }{\partial z} \right) 
+ \varepsilon(z) k_\parallel^2 \right] 
\widetilde{G}(\kv_\parallel, z,z') 
= 
4 \pi \delta(z-z').
\label{green-eq2}
\ee
The electrostatic Green function for the system depicted in Fig.~\ref{setup-slab}(b) for the case of two unit source charges with positions $z$ and $z'$ both inside the layered subspace $\ell/2 < z < \ell/2 + d$ can be found to be~(cf. Appendix~\ref{app:A}) 
\ba
&&\widetilde{G}(\kv_\parallel, z,z')
\label{green-multilayer}
\\
&\!\!=\!\!&
\frac{2\pi}{\varepsilon_2 k_\parallel} \, e^{- k_\parallel |z-z'|} 
- \frac{2\pi}{\varepsilon_2 k_\parallel} e^{k_\parallel (z - z')}
\nonumber\\
&&+
\frac{
2\pi 
\Theta_2
\big[
e^{- k_\parallel (z+z' - \ell)} 
+
R_{22'} \, e^{- k_\parallel (z - z' + 2d)}
\big]  
}
{\varepsilon_2 k_\parallel 
\left[
\Delta
 -
\Theta_2 R_{22'} \, e^{- 2 k_\parallel d)} 
\right]}
\nonumber\\
&&+
\frac{
2\pi 
\Delta 
\big[
e^{k_\parallel (z - z')} 
+
R_{22'} \, e^{k_\parallel (z + z' - \ell - 2d)}
\big]
}
{\varepsilon_2 k_\parallel 
\left[ 
\Delta 
- \Theta_2 R_{22'} \, e^{-2k_\parallel d)}  
\right]}.  
\nonumber
\ea
In the above, 
$\Delta \equiv 1 - R_{1m}R_{2m} e^{-2k_\parallel \ell}$,
 $\Theta_2 \equiv R_{2m}  - R_{1m} e^{-2k_\parallel \ell}$, 
 with the static reflection coefficients given by $R_{{\lambda }{\lambda' }} \equiv \frac{\varepsilon_{\lambda }-\varepsilon_{\lambda' }}{\varepsilon_{\lambda }+\varepsilon_{\lambda' }}$ and $\varepsilon_{\lambda }$ are the static dielectric permittivities of the ${\lambda }$th layer. 
The Green function for the system depicted in Fig.~\ref{setup-slab}(a) where $z, z' > \ell/2$ can be obtained from Eq.~(\ref{green-multilayer}) 
by taking the limit $d \to \infty$, and is given by 
\ba
&&\widetilde{G}(\kv_\parallel, z, z') 
= \frac{2\pi}{\varepsilon_2 k_\parallel} 
e^{- k_\parallel |z-z'|} 
\nonumber\\
&&\quad+ \frac{2\pi \big( R_{2m} e^{k_\parallel \ell} - R_{1m} e^{- k_\parallel \ell} \big)}{\varepsilon_2 k_\parallel \big( 1 - R_{1m} R_{2m} e^{- 2 k_\parallel \ell} \big)} 
e^{- k_\parallel (z+z')}. 
\label{G-direct}
\ea
As mentioned in the Introduction, we focus on the system being at zero temperature, which means that thermal fluctuations can be neglected. We will consider two types of dipolar disorder: (i)~{\it bulk disorder}, for which the local polarization is randomly distributed inside the bulk of a slab, and (ii)~{\it surface disorder}, where the local polarization is randomly distributed across the surface of a slab. 
For our material example, we will consider a relaxor ferroelectric in a disordered glassy phase, for which the local polarization is randomly distributed both spatially and orientationally. 
As the correlation length of the electric dipole fluctuations is set by the typical size of the polar nanoregion in a relaxor ferroelectric, we can assume that this correlation length is effectively zero if we consider slab surface (or atom-surface) separations much larger than the typical nanoregion size. 
If the bulk disorder occurs in layer 2, these considerations allow us to model the bulk disorder statistics by 
\ba
\overline{P_{{\lambda } i}(\rv)} &\!\!=\!\!& 0; 
\nonumber\\
\overline{P_{{\lambda } i}(\rv) P_{{\lambda' } j}(\rv')} 
&\!\!=\!\!& 
\delta_{ij} 
\delta(\rv - \rv')
\delta_{{\lambda } 2} \delta_{{\lambda' } 2} \Gamma_{B}.  
\label{stats-B}
\ea
The overline denotes an average performed over the quenched disorder statistics, and $\Gamma_{B}$ denotes the mean square quenched electric dipole moment per unit volume. 

For surface disorder, the variances of the quenched random dipoles in the directions parallel and perpendicular to the surface can be the same (orientational isotropy) or different (orientational anisotropy). 
If the surface disorder occurs on layer 2 adjacent to the gap medium as depicted in Fig.~\ref{setup-slab}, the surface disorder statistics for the case of orientational anisotropy can be modeled by
 \ba
\overline{P_{{\lambda } i}(\rv)} &\!\!=\!\!& 0; 
\nonumber\\
\overline{P_{{\lambda } i}(\rv) P_{{\lambda' } j}(\rv')} 
&\!\!=\!\!& 
\big( 
\delta_{ij}^{\parallel} \Gamma_{S}^{\parallel}
+ \delta_{i,z} \delta_{j,z} \Gamma_{S}^{\perp}
\big)
\delta(\rv_\parallel - \rv'_\parallel)
\delta_{{\lambda } 2} \delta_{{\lambda' } 2}
\nonumber\\
&&\times
 \delta(z-\ell/2) 
\delta(z'-\ell/2) 
. 
\label{stats-S-aniso}
\ea
Here, $\rv_\parallel = (x,y)$ and $\rv'_\parallel = (x',y')$ are vectors parallel to the plane of the surface, and $\delta_{ij}^{\parallel} = 1$ if $i=j=x$ or $i=j=y$ and $\delta_{ij}^{\parallel} = 0$ otherwise. The symbol $\Gamma_{S}^{\parallel}$ ($\Gamma_{S}^{\perp}$) denotes the mean square component of the quenched electric dipole moment per unit area parallel (perpendicular) to the surface. 
For the case of orientational isotropy, the surface disorder statistics becomes 
\ba
\overline{P_{{\lambda } i}(\rv)} &\!\!=\!\!& 0; 
\nonumber\\
\overline{P_{{\lambda } i}(\rv) P_{{\lambda' } j}(\rv')} 
&\!\!=\!\!& 
\delta_{ij}  \delta(\rv_\parallel - \rv'_\parallel)
\delta_{{\lambda } 2} \delta_{{\lambda' } 2}
\nonumber\\
&&\times
 \delta(z-\ell/2) 
\delta(z'-\ell/2) \Gamma_{S}, 
\label{stats-S}
\ea
where $\Gamma_{S}$ denotes the mean square quenched electric dipole moment per unit area.

\section{Fluctuation forces induced by bulk disorder}

Let us first consider the case where there are quenched random electric dipoles inside the bulk of the material (as opposed to being on the surface).

\subsection{Two single-layered slabs separated by a gap}

On averaging the electrostatic energy over the bulk polarization disorder using the statistics in Eq.~(\ref{stats-B}), the electrostatic energy of the quenched dipolar charges for the system shown in Fig.~\ref{setup-slab}(a) is given by 
\ba
\overline{U_E} 
&\!\!=\!\!&  
\frac{1}{2} \Gamma_{B} \! \int_{L2}\!d^3\rv \, {\bm\nabla}\cdot{\bm\nabla}' G(\rv,\rv') |_{\rv'\to\rv}.   
\label{energy-B}
\ea
Here, $\int_{L2}\!d^3\rv$ is a volume integral over the subspace of layer 2. 
The disorder-averaged energy is divergent and can be regularized by subtracting the contribution for $\ell\to\infty$, as we show in Appendix~\ref{app:reg}. 
We find that the regularized energy is given by   
\ba
U_{SS}^{(B)}
&\!\!=\!\!&  
\frac{\Gamma_{B} (1-R_{m2}^2) \, {{\rm Li}}_2(R_{m1} R_{m2})}
{8 \varepsilon_2 R_{m2} \ell^2} \mathcal{S}. 
\label{USSB}
\ea
In the above, ${\rm Li}_2(z)$ is the dilogarithm function and $\mathcal{S}$ is the cross-sectional area of the slab's surface. 
The normal component of the slab-slab fluctuation stress $\mathcal{P}_{SS}^{(B)} = - \frac{1}{\mathcal{S}} \frac{\partial U_{SS}^{(B)}}{\partial \ell}$ induced by quenched dipolar disorder in the bulk of the slabs is found to be 
\ba
\label{fSSB}
\mathcal{P}_{SS}^{(B)}
&\!\!=\!\!&  
\frac{\Gamma_{B} (1-R_{m2}^2) \, {{\rm Li}}_2(R_{m1} R_{m2})}
{4 R_{m2} \varepsilon_2 \ell^3}.
\ea
The result indicates that the bulk dipolar disorder-induced pressure can be attractive or repulsive, depending on the relative magnitudes of $\varepsilon_1$, $\varepsilon_m$ and $\varepsilon_2$. Correspondingly, $\mathcal{P}_{SS}^{(B)}$ can also be attractive or repulsive, depending on the relative magnitudes of the dielectric permittivities. 
For instance, if layer 1 is disorder-free and layer 2 contains quenched bulk disorder, $\mathcal{P}_{SS}^{(B)}$ is repulsive if $\varepsilon_1 < \varepsilon_m < \varepsilon_2$. Analogously, if layer 1 contains quenched bulk disorder and layer 2 is disorder-free, 
$\mathcal{P}_{SS}^{(B)}$ would also be repulsive if $\varepsilon_2 < \varepsilon_m < \varepsilon_1$. 
 This would be the case if layer 1 is KLT (with dielectric permittivity of the order of 1000), layer 2 is Si (with dielectric permittivity 11.74) and the intervening gap is filled with glycerol (with dielectric permittivity 42.4). The Casimir-Lifshitz pressure in the two-slab system is also repulsive if the dielectric contrast rule $\varepsilon_1(i\xi) \gtrless \varepsilon_m(i\xi) \gtrless \varepsilon_2(i\xi)$ (where $i\xi$ is the imaginary frequency) holds for the dominant frequency contributions (the ``Dzyaloshinskii-Lifshitz-Pitaevskii (DLP) condition"~\cite{esteso2016,DLP}). 
For a system that obeys the DLP condition and experiences the ``nanolevitation effect" in the absence of disorder, the said effect can be further enhanced if the system contains quenched dipolar disorder which obeys the static dielectric contrast rule $\varepsilon_1(0) \gtrless \varepsilon_m(0) \gtrless \varepsilon_2(0)$. One can regard this rule as a static analogue of the DLP condition.

\subsection{A bilayered slab and a single-layered slab separated by a gap}

The electrostatic energy of the quenched random polarizations for the system shown in Fig.~\ref{setup-slab}(b) is similar to Eq.~(\ref{energy-B}), with the Green function being now given by Eq.~(\ref{green-multilayer}) in Fourier space. 
The energy also involves an integral which is divergent and needs to be regularized, details of which are given in Appendix~\ref{app:reg}. We find that the regularized energy is given by 
\ba
\label{USSB'}
&&U_{SS}^{(B) \prime} 
=
\frac{\Gamma_B \mathcal{S}}{2 \varepsilon_2} \!
\int_0^\infty \!\!\!\!\! dk_\parallel \, k_\parallel 
\\
&&\quad\times 
\frac{(1 - e^{-2k_\parallel d}) (\Theta_2 - \Delta R_{2m}) (1 + R_{22'}^2 e^{- 2 k_\parallel d})}
{(\Delta - \Theta_2 R_{22'} \, e^{- 2 k_\parallel d}) 
(1 - R_{2m} R_{22'} \, e^{- 2 k_\parallel d})}.  
\nonumber
\ea
Here, layer 2 has a thickness $d$ and contains quenched disorder, whilst layers $1$ and $2'$ are both semi-infinite and disorder-free. 
From the above, the bulk dipolar disorder-induced pressure $\mathcal{P}_{SS}^{(B)}$ can be obtained using the formula
\be
\label{fSSB'}
\mathcal{P}_{SS}^{(B) \prime}
=
- \frac{1}{\mathcal{S}} \frac{\partial U_{SS}^{(B) \prime}}{\partial \ell}. 
\ee

\subsection{An atom above a single-layered slab}

\begin{figure}[h]
    \centering
      \includegraphics[width=0.48\textwidth]{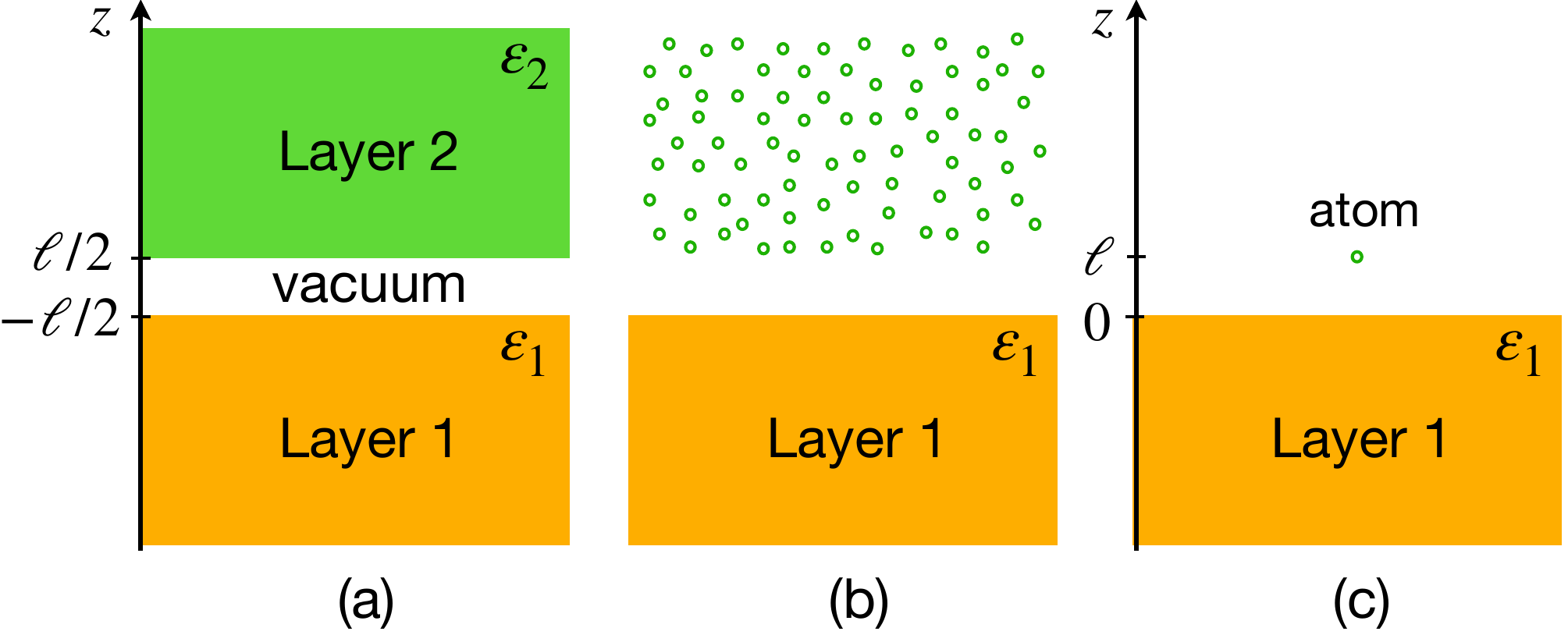}
      \caption{(a)~A disorder-free single-layered slab (layer 2) with dielectric permittivity $\varepsilon_2$ separated from a single-layered slab (layer 1) with dielectric permittivity $\varepsilon_1$ containing quenched random electric dipoles, with an intervening vacuum gap ($\varepsilon_m = 1$). 
      (b)~Rarefaction: the same system as in (a), with the slab in layer 2 replaced by a homogeneous and non-interacting gas of atoms. 
      (c)~A neutral atom in vacuum above a semi-infinite single-layered slab with dielectric permittivity $\varepsilon_1$. } 
      \label{setup-atom-singlelayer}
\end{figure} 

\begin{figure}[h]
    \centering
      \includegraphics[width=0.42\textwidth]{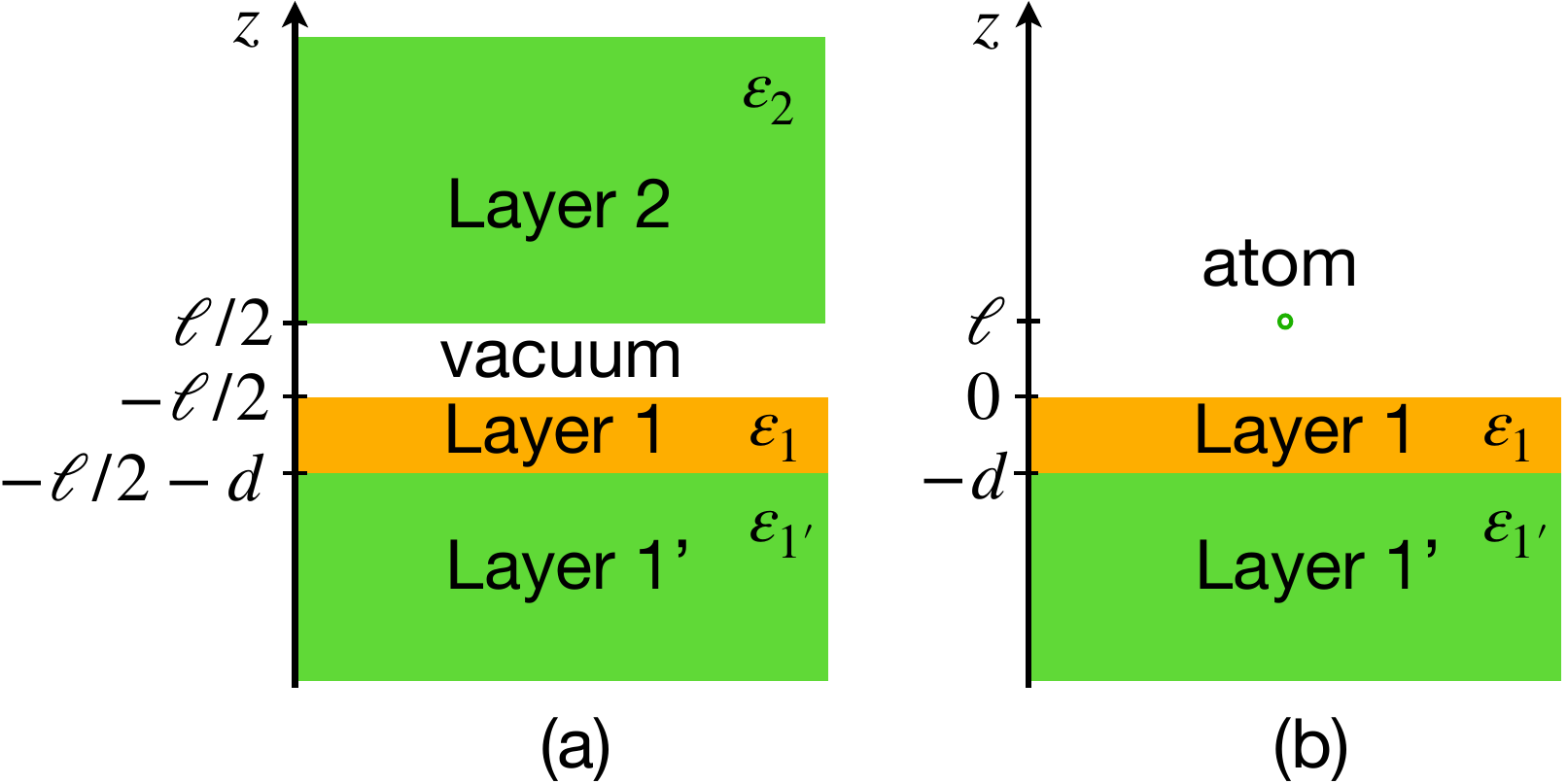}
      \caption{(a)~A pair of coplanar slabs separated by a vacuum gap ($\varepsilon_m = 1$). Layer 2 is single-layered and semi-infinite, with dielectric permittivity $\varepsilon_{2}$. Layer 1 has a thickness $d$ and dielectric permittivity $\varepsilon_1$, whilst layer $1'$ is semi-infinite and has a dielectric permittivity $\varepsilon_{1'}$. Layer 1 contains quenched dipolar disorder, whilst layers $1'$ and $2$ are disorder-free.
      (b)~A neutral atom in vacuum above a bilayered slab with the top layer having dielectric permittivity $\varepsilon_1$ and thickness $d$, and the bottom semi-infinite layer having dielectric permittivity $\varepsilon_{1'}$. This system can be obtained from that in (a) by rarefying layer 2, as explained in the text.} 
      \label{setup-atom-bilayer}
\end{figure} 

We next consider a system comprising an unpolarized ground-state atom in the vacuum above a semi-infinite dielectric slab which contains quenched random electric dipoles in the bulk (cf. Fig.~\ref{setup-atom-singlelayer}(c)). The disorder-induced  force on the atom arising from the quenched random bulk polarizations can be obtained via the rarefaction method~\cite{bordag2009,lifshitz1955,lu2025b}. Consider starting with the system shown in Fig.~\ref{setup-atom-singlelayer}(a), where the top slab (layer 2) is now disorder-free and the bottom slab (layer 1) contains quenched bulk dipolar disorder with a mean square electric dipole moment per unit volume $\Gamma_B$. The corresponding regularized disorder-averaged electrostatic energy of the quenched dipoles is given by Eq.~(\ref{USSB}) with the layer indices ``1" and ``2" interchanged, viz., 
\ba
U_{SS}^{(B)}
&\!\!=\!\!&  
\frac{\Gamma_{B} (1-R_{m1}^2) \, {{\rm Li}}_2(R_{m1} R_{m2})}
{8 \varepsilon_1 R_{m1} \ell^2} \mathcal{S}. 
\label{USSB1}
\ea
Next, we replace the dielectric slab occupying the subspace of layer 2 by a homogeneous, non-interacting gas of atoms whose number density is $n$ (cf. Fig.~\ref{setup-atom-singlelayer}(b)). Operationally, this means approximating $\varepsilon_2 \approx 1 + 4\pi n \alpha_0$ (where $\alpha_0$ is the atomic polarizability and $n\alpha_0 \ll 1$) and $R_{m2} \approx -2\pi n \alpha_0$ in Eq.~(\ref{USSB1}). The energy becomes 
\ba
U_{AS}^{(B)}
&\!\!=\!\!& 
- \frac{\pi \Gamma_{B} n\alpha_0}
{(\varepsilon_1+1)^2 \ell^2} 
\, \mathcal{S}. 
\label{UASB-single}
\ea
In the notation for the energy $U_{AS}^{(B)}$, the subscripts ``AS" remind us that this is the energy of the slab-atom gas system, and the superscript ``(B)" indicates that the quenched disorder occurs in the bulk. 
$U_{AS}^{(B)}$ can also be regarded as the change in the energy of a non-interacting gas of atoms induced by the presence of the quenched polarization disorder.  
We can thus equate this quantity to the sum of the energy changes $\delta E_{{\rm atom}}^{(B)}$ of individual atoms induced by the quenched polarization disorder: 
\be
U_{AS}^{(B)} = n \mathcal{S} \int_\ell^\infty \!\!\! dz \, 
\delta E_{{\rm atom}}^{(B)}(z),  
\label{UASB}
\ee
where $z$ is the position of a given atom in the gas occupying the subspace of layer 2. 
We can invert the expression to obtain the bulk polarization disorder-induced energy change of an atom located at a distance $\ell$ from the surface of layer 1 (cf. Fig.~\ref{setup-atom-singlelayer}(c)): 
\be
\delta E_{{\rm atom}}^{(B)}(\ell)
=
- \frac{1}{n \mathcal{S}} \frac{\partial U_{AS}^{(B)}}{\partial \ell}.  
\label{deltaEatom}
\ee
From Eq.~(\ref{UASB}) and the equation above, we find   
\be
\delta E_{{\rm atom}}^{(B)}(\ell)
=
- \frac{2\pi \Gamma_{B} \alpha_0}
{(\varepsilon_1+1)^2 \ell^3}.   
\ee
The bulk polarization disorder-induced force acting on the atom is then 
\be
f_{AS}^{(B)} = - \frac{\partial \delta E_{{\rm atom}}^{(B)}}{\partial \ell} 
= - \frac{6\pi \Gamma_{B} \alpha_0}
{(\varepsilon_1+1)^2 \ell^4}.  
\label{fAS-singlelayer}
\ee
The force calculated using the above formula is in units of dynes. To convert to Newtons, we have to multiply the result by a factor of $10^{-5}$. 

\subsection{An atom above a bilayered slab}

In a similar vein, we can apply the rarefaction method to obtain the bulk polarization disorder-induced force acting on an atom in the vacuum above a bilayered slab. We consider the system shown in Fig.~\ref{setup-atom-bilayer}(a), in which layers $1'$ and $2$ are disorder-free, and layer 1 contains quenched bulk dipolar disorder with a mean square dipole moment per unit volume $\Gamma_B$. 
For this system, instead of Eq.~(\ref{USSB'}), the regularized electrostatic energy is given by 
\ba
&&U_{SS}^{(B) \prime} 
= 
\frac{\Gamma_B \mathcal{S}}{2 \varepsilon_1} \!
\int_0^\infty \!\!\!\!\! dk_\parallel \, k_\parallel 
\\
&&\quad\times 
\frac{(1 - e^{-2k_\parallel d}) (\Theta_1 - \Delta R_{1m}) (1 + R_{11'}^2 e^{- 2 k_\parallel d})}
{(\Delta - \Theta_1 R_{11'} \, e^{- 2 k_\parallel d}) 
(1 - R_{1m} R_{11'} \, e^{- 2 k_\parallel d})}, 
\nonumber
\ea
where $\Theta_1 \equiv R_{1m} - R_{2m} e^{-2k_\parallel \ell}$. 
We next replace the slab in layer 2 by a homogeneous gas of non-interacting atoms each of polarizability $\alpha_0$ and number density $n$. To leading order in $n\alpha_0$, this implies that $\varepsilon_2 \approx 1 + 4\pi n \alpha_0$ and $R_{2m} \approx 2\pi n \alpha_0$. The energy $U_{SS}^{(B) \prime}$ becomes 
\ba
\label{UASB'}
&&U_{AS}^{(B) \prime} 
=
- \frac{4 \pi n \alpha_0 \Gamma_B \mathcal{S}}{(\varepsilon_1+1)^2}
\\
&&\times 
\int_0^\infty \!\!\!\!\! dk_\parallel \, 
\frac{k_\parallel \, e^{-2k_\parallel\ell} (1 + R_{11'}^2 e^{- 2 k_\parallel d}) (1 - e^{-2k_\parallel d}) }
{(1 - R_{1m}R_{11'} \, e^{- 2 k_\parallel d})^2}.
\nonumber
\ea
The change in the energy of an individual atom $\delta E_{{\rm atom}}^{(B) \prime}(\ell) = - \frac{1}{n \mathcal{S}} \frac{\partial U_{AS}^{(B) \prime}}{\partial \ell}$ is found to be 
\ba
\label{EB'}
&&\delta E_{{\rm atom}}^{(B) \prime}(\ell)
=
- \frac{8 \pi \alpha_0 \Gamma_B}{(\varepsilon_1+1)^2}
\\
&&\times 
\int_0^\infty \!\!\!\!\! dk_\parallel \, 
\frac{k_\parallel^2 \, e^{-2k_\parallel\ell} (1 + R_{11'}^2 e^{- 2 k_\parallel d}) (1 - e^{-2k_\parallel d}) }
{(1 - R_{1m}R_{11'} \, e^{- 2 k_\parallel d})^2}.
\nonumber
\ea
We find the bulk dipolar disorder-induced force acting on the atom, $f_{AS}^{(B) \prime}  = - \frac{\partial \delta E_{{\rm atom}}^{(B) \prime}}{\partial \ell}$: 
\ba
\label{fAS'}
&&f_{AS}^{(B) \prime} 
=
- \frac{16 \pi \alpha_0 \Gamma_B}{(\varepsilon_1+1)^2}
\\
&&\times 
\int_0^\infty \!\!\!\!\! dk_\parallel \, 
\frac{k_\parallel^3 \, e^{-2k_\parallel\ell} (1 + R_{11'}^2 e^{- 2 k_\parallel d}) (1 - e^{-2k_\parallel d}) }
{(1 - R_{1m}R_{11'} \, e^{- 2 k_\parallel d})^2}.
\nonumber
\ea
This becomes Eq.~(\ref{fAS-singlelayer}) in the limit $d \to \infty$, as we expect.

\section{Fluctuation forces induced by surface disorder}

Next, we consider the disorder-induced forces arising from quenched random surface dipoles. For the case where the surface disorder is orientationally isotropic, we can average the electrostatic energy using the statistics given in Eq.~(\ref{stats-S}). 
This yields
\ba
\label{UE-surface}
&&\overline{U_E} 
\label{energy-S}\\
&\!\!=\!\!&  
\frac{1}{2} \Gamma_{S} \! \int_{L2} \!d^2\rv_\parallel \, {\bm\nabla}\cdot{\bm\nabla}' G(\rv_\parallel,z; \rv'_\parallel,z') |_{\rv'_\parallel\to\rv_\parallel; z,z'\to \ell/2}.  
\nonumber
\ea
In the above, $\int_{L2}\!d^2\rv_\parallel$ represents an integral over the surface of layer 2. 
For the case where the surface disorder is orientationally anisotropic, the averaging is performed using Equation~(\ref{stats-S-aniso}), which yields 
\ba
&&\overline{U_E} 
\label{energy-S-aniso}\\
&\!\!=\!\!&  
\frac{1}{2} \Gamma_{S}^\parallel \! 
\int_{L2} \!d^2\rv_\parallel \, {\bm\nabla}_\parallel \!\cdot\! {\bm\nabla}'_\parallel 
G(\rv_\parallel,z; \rv'_\parallel,z') |_{\rv'_\parallel\to\rv_\parallel; z,z'\to \ell/2}
\nonumber\\
&&+ 
\frac{1}{2} \Gamma_{S}^\perp \! 
\int_{L2} \!d^2\rv_\parallel \, \frac{\partial}{\partial z} \frac{\partial}{\partial z'}  
G(\rv_\parallel,z; \rv'_\parallel,z') |_{\rv'_\parallel\to\rv_\parallel; z,z'\to \ell/2}.  
\nonumber
\ea
Here, ${\bm\nabla}_\parallel$ is the gradient operator in the $xy$ plane.  
Analogous to what we have done in the previous section, we will obtain expressions for the surface dipolar disorder-induced pressure between the slabs in the systems depicted by Fig.~\ref{setup-slab}, as well as obtain expressions for the surface dipolar disorder-induced force on an atom in the systems depicted by Figs.~\ref{setup-atom-singlelayer}(c) and \ref{setup-atom-bilayer}(b). 

\subsection{Two single-layered slabs separated by a gap}

We first consider the system depicted by Fig.~\ref{setup-slab}(a) with layer 1 being disorder-free and layer 2 having quenched surface dipolar disorder. The disorder-averaged electrostatic energy of the quenched surface electric dipoles, 
which is Eq.~(\ref{UE-surface}) for orientationally isotropic surface disorder and Eq.~(\ref{energy-S-aniso}) for orientationally anisotropic surface disorder, 
is divergent, and can be regularized by subtracting the contribution for $\ell\to\infty$.
After regularization (cf. Appendix~\ref{app:reg}), we find that the energy of the quenched surface electric dipoles is given by 
\ba
U_{SS}^{(S)}
&\!\!=\!\!&  
\frac{\Gamma_S^\ast(1-R_{m2}^2) \, {\rm Li}_3(R_{m1}R_{m2})}
{4 \varepsilon_2 R_{m2} \ell^3} \mathcal{S}. 
\label{USSS}
\ea
Here ${\rm Li}_3(z)$ is the trilogarithm function, and $\Gamma_S^\ast = \Gamma_S$ if the surface disorder is orientationally isotropic with disorder statistics given by Eq.~(\ref{stats-S}), and $\Gamma_S^\ast = \frac{1}{2}(\Gamma_S^\parallel + \Gamma_S^\perp)$ if the surface disorder is orientationally anisotropic with disorder statistics given by Eq.~(\ref{stats-S-aniso}). 
The slab-slab disorder-induced  pressure $\mathcal{P}_{SS}^{(S)}$ induced by quenched surface dipolar disorder is then 
\ba
\mathcal{P}_{SS}^{(S)}
&\!\!=\!\!&  
\frac{3 \Gamma_S^\ast (1-R_{m2}^2) \, {\rm Li}_3(R_{m1}R_{m2})}
{4 \varepsilon_2 R_{m2} \ell^4}. 
\label{fSSS}
\ea
The surface dipolar disorder-induced pressure decays with $\ell^{-4}$, which is stronger compared with the $\ell^{-3}$ power-law decay exhibited by the bulk dipolar disorder-induced pressure between two single-layered slabs. It can be attractive or repulsive, depending on the relative magnitudes of $\varepsilon_1$, $\varepsilon_m$ and $\varepsilon_2$. 
For example, $\mathcal{P}_{SS}^{(S)}$ is repulsive if $\varepsilon_1 < \varepsilon_m < \varepsilon_2$.

\subsection{A bilayered slab and a single-layered slab separated by a gap}

The electrostatic energy of quenched random surface electric dipoles for the system depicted in Figure~\ref{setup-slab}(b) can be calculated with the Green function in Eq.~(\ref{green-multilayer}). 
After performing the disorder average using Eq.~(\ref{stats-S}) if the surface disorder is orientationally isotropic or using Eq.~(\ref{stats-S-aniso}) if the surface disorder is orientationally anisotropic, and regularizing the integrals, we find (cf. Appendix~\ref{app:reg}) that the electrostatic energy is given by 
\ba
\label{USSS'}
&&U_{SS}^{(S) \prime}
\\
&\!\!=\!\!&  
\frac{\Gamma_S^\ast \mathcal{S}}{\varepsilon_2} \!  
\int_0^\infty \!\!\!\!\! dk_\parallel \, 
\frac{k_\parallel^2 \, (\Theta_2 - \Delta R_{2m}) (1 + R_{22'}^2 \, e^{- 4 k_\parallel d})}
{(\Delta - \Theta_2 R_{22'} \, e^{- 2 k_\parallel d}) (1 - R_{2m}R_{22'} \, e^{- 2 k_\parallel d})}. 
\nonumber
\ea
The corresponding disorder-induced  pressure arising from the quenched random surface dipoles can be obtained via  
\be
\mathcal{P}_{SS}^{(S) \prime}
=
- \frac{1}{\mathcal{S}} \frac{\partial U_{SS}^{(S)\prime}}{\partial \ell}. 
\label{PSSS'}
\ee

\subsection{An atom above a single-layered slab}

The case of an unpolarized ground-state atom above a semi-infinite dielectric slab containing quenched surface dipolar disorder is depicted in Fig.~\ref{setup-atom-singlelayer}(c), where layer 1 contains quenched surface dipolar disorder with a mean square electric dipole moment per unit area $\Gamma_S$.  As before, we can obtain the disorder-induced force on the atom by rarefying the slab in layer 2 depicted in Fig.~\ref{setup-atom-singlelayer}(a) into a homogeneous non-interacting gas of atoms of polarizability $\alpha_0$ and number density $n$. 
The disorder-averaged regularized energy of the slab-atom gas system is given by 
\be
U_{AS}^{(S)}
=
- \frac{2\pi n\alpha_0 \Gamma_{S}^\ast}
{(\varepsilon_1+1)^2 \ell^3} 
\, \mathcal{S}. 
\ee
In the notation above, the superscript ``(S)" refers to the quenched surface dipolar disorder. 
For a non-interacting gas of atoms, the energy $U_{AS}^{(S)}$ is also equal to the sum of the surface dipolar disorder-induced energy changes $\delta E_{{\rm atom}}^{(S)}$ of individual atoms in the gas: 
\be
U_{AS}^{(S)} = n \mathcal{S} \int_\ell^\infty \!\!\! dz \, 
\delta E_{{\rm atom}}^{(S)}(z). 
\ee
Inverting the expression leads to the surface dipolar disorder-induced energy change of an atom located at a distance $\ell$ from the surface of slab 1: 
\be
\delta E_{{\rm atom}}^{(S)}(\ell)
=
- \frac{1}{n \mathcal{S}} \frac{\partial U_{AS}^{(S)}}{\partial \ell}
=
- \frac{6\pi \Gamma_S^\ast \alpha_0}
{(\varepsilon_1+1)^2 \ell^4}.   
\ee
The surface dipolar disorder-induced force acting on the atom is then 
\be
f_{AS}^{(S)} = - \frac{\partial \delta E_{{\rm atom}}^{(S)}}{\partial \ell} 
= - \frac{24\pi \Gamma_S^\ast \alpha_0}
{(\varepsilon_1+1)^2 \ell^5}.  
\label{fASS}
\ee

\subsection{An atom above a bilayered slab}

We lastly consider the system depicted in Fig.~\ref{setup-atom-bilayer}(b), which comprises an atom above a bilayered slab consisting of a top layer of thickness $d$ and dielectric permittivity $\varepsilon_{1}$ and a semi-infinite bottom layer of dielectric permittivity $\varepsilon_{1'}$. The surface dipolar disorder 
occurs on the top surface of the slab. 
To obtain the surface dipolar disorder-induced force acting on the atom, 
we begin with the system shown in Fig.~\ref{setup-atom-bilayer}(a), for which the energy $U_{SS}^{(S) \prime}$ is obtained from Eq.~(\ref{USSS'}) by changing the layer indices $1 \to 2$, $2 \to 1$ and $2' \to 1'$:
\ba
&&U_{SS}^{(S) \prime}
\\
&\!\!=\!\!&  
\frac{\Gamma_S^\ast \mathcal{S}}{\varepsilon_1} \!  
\int_0^\infty \!\!\!\!\! dk_\parallel \, 
\frac{k_\parallel^2 \, (\Theta_1 - \Delta R_{1m}) (1 + R_{11'}^2 \, e^{- 4 k_\parallel d})}
{(\Delta - \Theta_1 R_{11'} \, e^{- 2 k_\parallel d}) (1 - R_{1m}R_{11'} \, e^{- 2 k_\parallel d})}. 
\nonumber
\ea
On rarefying the slab in the subspace of layer 2, the above energy becomes $U_{AS}^{(S) \prime}$, given by 
\be
U_{AS}^{(S) \prime}
=
- \frac{8\pi n \alpha_0 \Gamma_S^\ast \mathcal{S}}{(\varepsilon_1 + 1)^2} 
\int_0^\infty \!\!\!\!\! dk_\parallel \, 
\frac{k_\parallel^2 \, e^{-2k_\parallel \ell} (1 + R_{11'}^2 \, e^{- 4 k_\parallel d})}
{(1 - R_{1m}R_{11'} \, e^{- 2 k_\parallel d})^2}.
\ee
The change in the energy of an individual atom is obtained from the formula $\delta E_{{\rm atom}}^{(S) \prime} = - \frac{1}{n \mathcal{S}} \frac{\partial U_{AS}^{(S) \prime}}{\partial \ell}$, yielding
\be
\delta E_{{\rm atom}}^{(S) \prime}(\ell)
= 
- \frac{16\pi \alpha_0 \Gamma_S^\ast}{(\varepsilon_1 + 1)^2} 
\int_0^\infty \!\!\!\!\! dk_\parallel \, 
\frac{k_\parallel^3 \, e^{-2k_\parallel \ell} (1 + R_{11'}^2 \, e^{- 4 k_\parallel d})}
{(1 - R_{1m}R_{11'} \, e^{- 2 k_\parallel d})^2}.
\ee
We find the surface dipolar disorder-induced force acting on the atom via $f_{AS}^{(S) \prime}  = - \frac{\partial \delta E_{{\rm atom}}^{(S) \prime}}{\partial \ell} $: 
\be
f_{AS}^{(S) \prime} 
=
- \frac{32\pi \alpha_0 \Gamma_S^\ast}{(\varepsilon_1 + 1)^2} 
\int_0^\infty \!\!\!\!\! dk_\parallel \, 
\frac{k_\parallel^4 \, e^{-2k_\parallel \ell} (1 + R_{11'}^2 \, e^{- 4 k_\parallel d})}
{(1 - R_{1m}R_{11'} \, e^{- 2 k_\parallel d})^2}.
\label{fASS'}
\ee
For $d \to \infty$, the above becomes Eq.~(\ref{fASS}), as it should.

\section{Results and Discussion} 

\begin{figure}[h]
    \centering
      \includegraphics[width=0.48\textwidth]{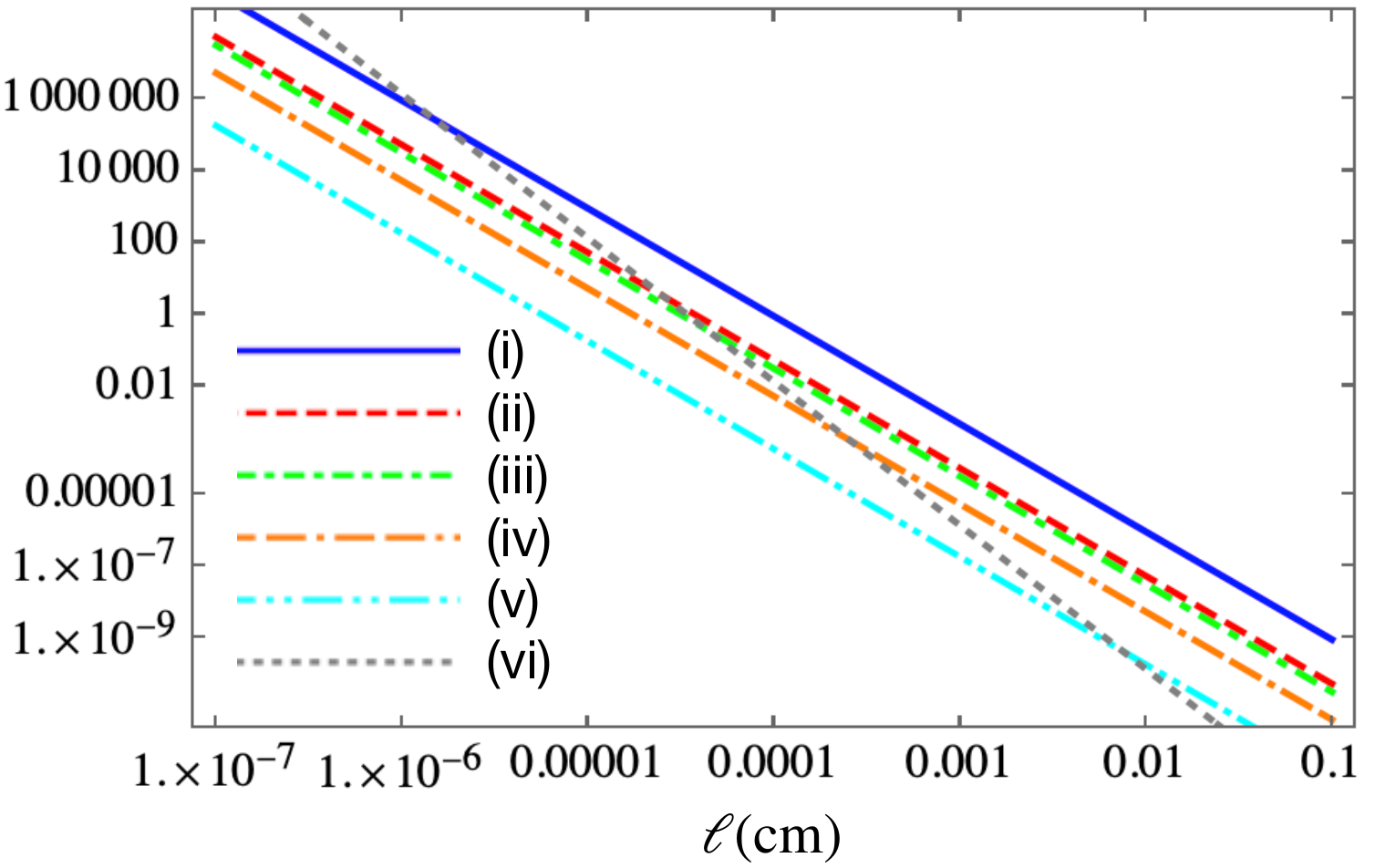}
      \caption{Behavior of the disorder-induced  pressure (in $\rm{dyn}/cm^2$, 
       vertical axis) as a function of gap width $\ell$ (in cm, horizontal axis) for the system shown in Fig.~\ref{setup-slab}(a), with layer 1 and the gap disorder-free and layer 2 having a mean square quenched dipole moment per unit volume $\Gamma_B$: 
      (i)~$\mathcal{P}_{SS}^{(B)}$ with $\varepsilon_1 = 11.74$, $\varepsilon_m = 42.4$, $\varepsilon_2 = 1000$ and $\Gamma_{B} = 4.4\times 10^{-8} \, {\rm esu^2/cm}$  (blue); 
      (ii)~$- \mathcal{P}_{SS}^{(B)}$ with $\varepsilon_1 = 11.74$, $\varepsilon_m = 1$, $\varepsilon_2 = 1000$ and $\Gamma_{B} = 4.4\times 10^{-8} \, {\rm esu^2/cm}$  (red dashed); 
      (iii)~$- \mathcal{P}_{SS}^{(B)}$ with $\varepsilon_1 = 3.7$, $\varepsilon_m = 1$, $\varepsilon_2 = 1000$ and $\Gamma_{B} = 4.4\times 10^{-8} \, {\rm esu^2/cm}$  (green dot-dashed); 
      (iv)~$- \mathcal{P}_{SS}^{(B)}$ with $\varepsilon_1 = 11.74$, $\varepsilon_m = 1$, $\varepsilon_2 = 1000$ and $\Gamma_{B} = 4.4\times 10^{-9} \, {\rm esu^2/cm}$  (orange dot-dashed-dashed); 
      (v)~$- \mathcal{P}_{SS}^{(B)}$ with $\varepsilon_1 = 11.74$, $\varepsilon_m = 1$, $\varepsilon_2 = 1000$ and $\Gamma_{B} = 1.5\times 10^{-10} \, {\rm esu^2/cm}$  (cyan dot-dot-dashed); 
      (vi)~the Casimir pressure magnitude $-\mathcal{P}_{\rm{Cas}}$ between two ideal conductors with a vacuum gap (gray dotted). 
 As the point dipole approximation works well for $\ell$ larger than the typical polar nanoregion size, the small $\ell$ region is an extrapolation.} 
      \label{single-layer}
\end{figure} 

\begin{figure}[h]
    \centering
      \includegraphics[width=0.48\textwidth]{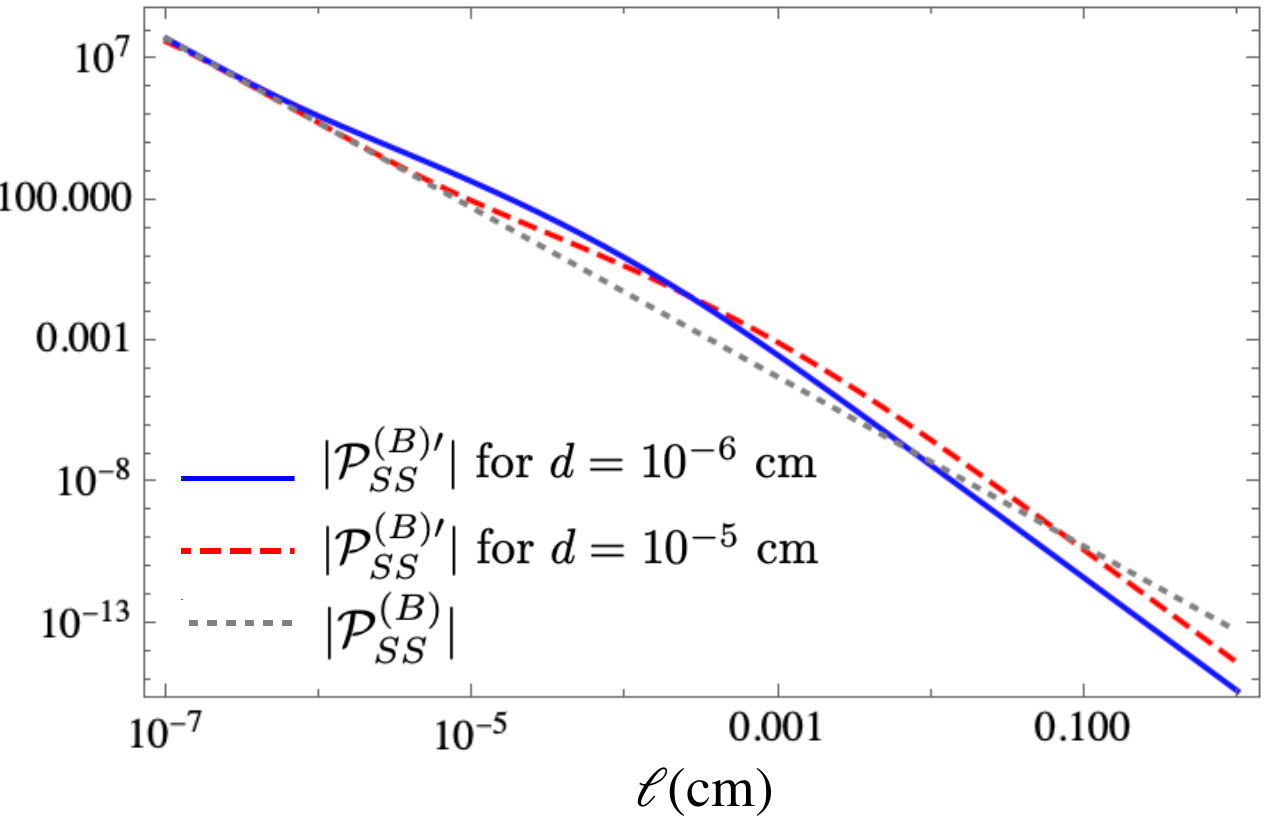}
      \caption{Behavior of the disorder-induced  pressure magnitude ($\rm{dyn}/cm^2$) 
      as a function of the vacuum gap width $\ell$ (cm): 
      (i)~$|\mathcal{P}_{SS}^{(B) \prime}|$ for the system shown in Fig.~\ref{setup-slab}(b) with the gap being the vacuum, $\varepsilon_{1} = \varepsilon_{2'} = 11.74$, $\varepsilon_2 = 1000$, layers 1 and $2'$ are disorder-free, and layer 2 has a mean square quenched dipole moment per unit volume $\Gamma_{B} = 4.4\times 10^{-8} \, {\rm esu^2/cm}$ and a thickness $d = 10^{-6}$ cm (blue curve); 
      (ii)~$|\mathcal{P}_{SS}^{(B) \prime}|$ for the same system as in (i) but with $d = 10^{-5}$ cm, all other parameter values being the same (red dashed curve);
      (iii)~$|\mathcal{P}_{SS}^{(B)}|$ for the system shown in Fig.~\ref{setup-slab}(a) with $\varepsilon_1 = 11.74$, $\varepsilon_m = 1$, $\varepsilon_2 = 1000$ and $\Gamma_{B} = 4.4\times 10^{-8} \, {\rm esu^2/cm}$ (gray dotted curve).  As the point dipole approximation works well for $\ell$ larger than the typical polar nanoregion size, the small $\ell$ region is an extrapolation.} 
      \label{bilayer}
\end{figure} 
In this section, we will apply some of the formulas we derived in the previous sections to a few examples, looking in particular at the case of quenched bulk dipolar disorder.
Rather than focus on a specific material, we consider the effect of varying the dielectric permittivity and the disorder parameter $\Gamma_B$ on the behavior of the bulk dipolar disorder-induced pressure and force behavior. This is motivated, firstly, by the wide range of possible values for the zero-temperature static dielectric permittivity of relaxor ferroelectric materials (which can be of the order of $10^2$ for ${\rm KTaO_3}$-based relaxor ferroelectrics~\cite{cai2015} to more than $2\times 10^4$ for ${\rm SrTiO_3}$-based relaxor ferroelectrics~\cite{muller1979}). Secondly, even for the same material, different samples may possess different values for the zero-temperature static dielectric permittivity~\cite{salce1994}. Thirdly, the mean square polarization appears to have been measured only for a few material systems (viz., PLZT~\cite{burns1983b} and PMN~\cite{cross1987}), with the measurements being done at nonzero temperatures. For our analysis, we will assume that the zero-temperature relaxor ferroelectric system is in the disordered glassy state. The relevant phenomenological parameters that influence the strength of the dipolar disorder-induced force are the dielectric permittivity, the mean square polarization and the typical size of the polar nanoregions. 
The idea is that by using the formulas we have derived, the experimentalist would be able to determine for a particular system and range of separations whether or not the quenched dipolar disorder-induced force can be neglected relative to the Casimir-Lifshitz force by measuring the values of the phenomenological parameters and calculating the corresponding force magnitudes. 

\

 As our first example, we consider the system in Fig.~\ref{setup-slab}(a). 
 Layer 1 is disorder-free, whilst layer 2 contains only quenched bulk dipolar disorder. 
In Fig.~\ref{single-layer}, we plot the behavior of the quenched bulk disorder-induced  pressure $\mathcal{P}_{SS}^{(B)}$ as a function of the gap width $\ell$. 
 The disorder-induced  pressure behaviors have been plotted based on Eq.~(\ref{fSSB}).  From the behavior of curve~(i), we see that the disorder-induced  pressure is repulsive for $\varepsilon_1 = 11.74$, $\varepsilon_m = 42.4$ and $\varepsilon_2 = 1000$ (these numbers correspond respectively to the static dielectric constants of Si, glycerol and KLT). 
This is related to the factor ${\rm Li}_2(R_{m1}R_{m2})/R_{m2}$ in Eq.~(\ref{fSSB}), which is positive if $\varepsilon_1 < \varepsilon_m < \varepsilon_2$.  
On the other hand, the disorder-induced  pressure is attractive if the intervening gap is the vacuum, as we see from the behavior of curves~(ii), (iii), (iv) and (v). The disorder-induced  pressure becomes less attractive if the dielectric permittivity $\varepsilon_1$ is reduced whilst keeping the other parameters the same, as we see on going from curve (ii) to curve (iii), for which $\varepsilon_1$ decreases from 11.74 to 3.7 (the latter value corresponding to the static dielectric constant of $\rm{SiO_2}$). 
Lastly, we can compare the strength of the bulk disorder-induced pressure with the zero-temperature Casimir pressure $\mathcal{P}_{\rm{Cas}}$ between two ideal conductors separated by a vacuum gap:   
\be
\mathcal{P}_{\rm{Cas}} = - \frac{\pi^2 \hbar c}{240 \, \ell^4}.  
\label{f-Casimir}
\ee
Here, $\hbar = 1.055\times 10^{-27} \, {\rm ergs.s}$ and $c = 2.998\times 10^{10} \, {\rm cm/s}$. The behavior of $\mathcal{P}_{\rm{Cas}}$ is represented by curve~(vi). We make the comparison with $\mathcal{P}_{\rm{Cas}}$ in the ideal conductor limit as it provides the theoretical upper bound on the strength of Casimir-Lifshitz attraction between two dielectric bodies separated by a vacuum gap. 
From curves (ii), (iv) and (v), we see that the quenched bulk disorder-induced pressure can be more attractive than $\mathcal{P}_{\rm{Cas}}$ and would need to be taken into account if $\ell$ is large enough. 
As the typical polar nanoregion (PNR) size can be as small as 1.5 nm or as large as 18 nm, there will be a corresponding range of possible values for the mean square quenched dipole moment per unit volume $\Gamma_B$. 
For example, if we take the typical PNR size to be 10 nm and take a representative value of the root-mean-square (rms) polarization as $2.1\times 10^5 \, {\rm esu. cm^{-2}}$~\cite{burns1983b}, $\Gamma_B = 4.4\times 10^{-8}$ $\rm{esu^2/cm}$. The corresponding disorder-induced pressure behavior is shown as curve (ii), where we see that $\mathcal{P}_{SS}^{(B)}$ is more attractive than $\mathcal{P}_{\rm{Cas}}$ for gap widths larger than 
$\ell_0 = 2.55\times10^{-5}$ cm. 
The value $\ell_0$ is larger if $\Gamma_B$ is smaller: for $\Gamma_B = 4.4\times 10^{-9}$ $\rm{esu^2/cm}$ (curve (iv)), 
$\ell_0 = 2.55\times10^{-4}$ cm. 
If we set the typical PNR size to be at the lower limit of 1.5 nm with the rms polarization value of $2.1\times 10^5 \, {\rm esu. cm^{-2}}$, $\Gamma_B = 1.5\times 10^{-10}$ $\rm{esu^2/cm}$, and $\ell_0 = 7.45 \times 10^{-3}$ cm. 
These crossover values provide a conservative benchmark, being based on a comparison with the Casimir pressure for ideal conductors. 

\

In Fig.~\ref{bilayer}, we plot the behavior of the quenched bulk disorder-induced pressure $\mathcal{P}_{SS}^{(B) \prime}$ for the system depicted in Fig.~\ref{setup-slab}(b), which consists of a bilayered top slab and a single-layered bottom slab separated by a vacuum gap. Layers 1 and $2'$ are disorder-free and have the same dielectric permittivity values, viz.,  $\varepsilon_1 = \varepsilon_{2'} = 11.74$, whilst for layer 2 $\Gamma_{B} = 4.4\times 10^{-8} \, {\rm esu^2/cm}$ and $\varepsilon_2 = 1000$. 
The behavior of $\mathcal{P}_{SS}^{(B) \prime}$ is plotted using Eq.~(\ref{fSSB'}), and is represented by the blue curve for $d = 10^{-6}$ cm and the red dashed curve for $d = 10^{-5}$ cm.  
For comparison, we have also plotted the disorder-induced  pressure $\mathcal{P}_{SS}^{(B)}$ between two coplanar semi-infinite single-layered slabs separated by a vacuum gap, with $\varepsilon_1 = 11.74$, $\varepsilon_m = 1$ and $\Gamma_{B} = 4.4\times 10^{-8} \, {\rm esu^2/cm}$ and $\varepsilon_2 = 1000$ (gray dotted curve -- this is the same curve as curve (ii) from Fig.~\ref{single-layer}). 

\

We first note that for $\ell$ smaller than $d$, the behavior of $\mathcal{P}_{SS}^{(B) \prime}$ approaches that of $\mathcal{P}_{SS}^{(B)}$, with a $\ell^{-3}$ scaling behavior 
(i.e., the blue and red dashed curves converge onto the gray dotted curve). 
This is because at sufficiently short separations, layer 2 effectively appears to be infinitely thick, and we recover the disorder-induced  pressure behavior for the system depicted in Fig.~\ref{setup-slab}(a). 
Secondly, as we show in Appendix~\ref{app:C}, for $\ell \gg d$, the power-law scaling behavior of the {\em bulk} disorder-induced pressure $\mathcal{P}_{SS}^{(B) \prime}$ changes to $\ell^{-4}$, which is the scaling behavior of the {\em surface} disorder-induced pressure between two coplanar semi-infinite single-layered slabs (cf. Eq.~(\ref{fSSS})). This is because for distances much greater than $d$, the bulk dipolar disorder in layer 2 effectively appears to be localized on the surface of layer $2'$. 
Thirdly, we note that the presence of an additional dielectric interface in the system in Fig.~\ref{setup-slab}(b) compared to that in Fig.~\ref{setup-slab}(a) gives rise to additional image dipole contributions, which can make $\mathcal{P}_{SS}^{(B) \prime}$ more attractive than $\mathcal{P}_{SS}^{(B)}$ for sufficiently small gap widths, as we see in Fig.~\ref{bilayer}. 
The range of separations over which $\mathcal{P}_{SS}^{(B) \prime}$ is more attractive increases (decreases) if $d$ increases (decreases). 
From Fig.~\ref{bilayer}, we see that for $d = 10^{-6}$ cm, $\mathcal{P}_{SS}^{(B) \prime}$ is more attractive than $\mathcal{P}_{SS}^{(B)}$ for $\ell < 7.35 \times 10^{-3}$ cm, whilst for $d = 10^{-5}$ cm, $\mathcal{P}_{SS}^{(B) \prime}$ is more attractive than $\mathcal{P}_{SS}^{(B)}$ if $\ell < 7.35 \times 10^{-2}$ cm.  

\ 

\begin{figure}[h]
    \centering
      \includegraphics[width=0.48\textwidth]{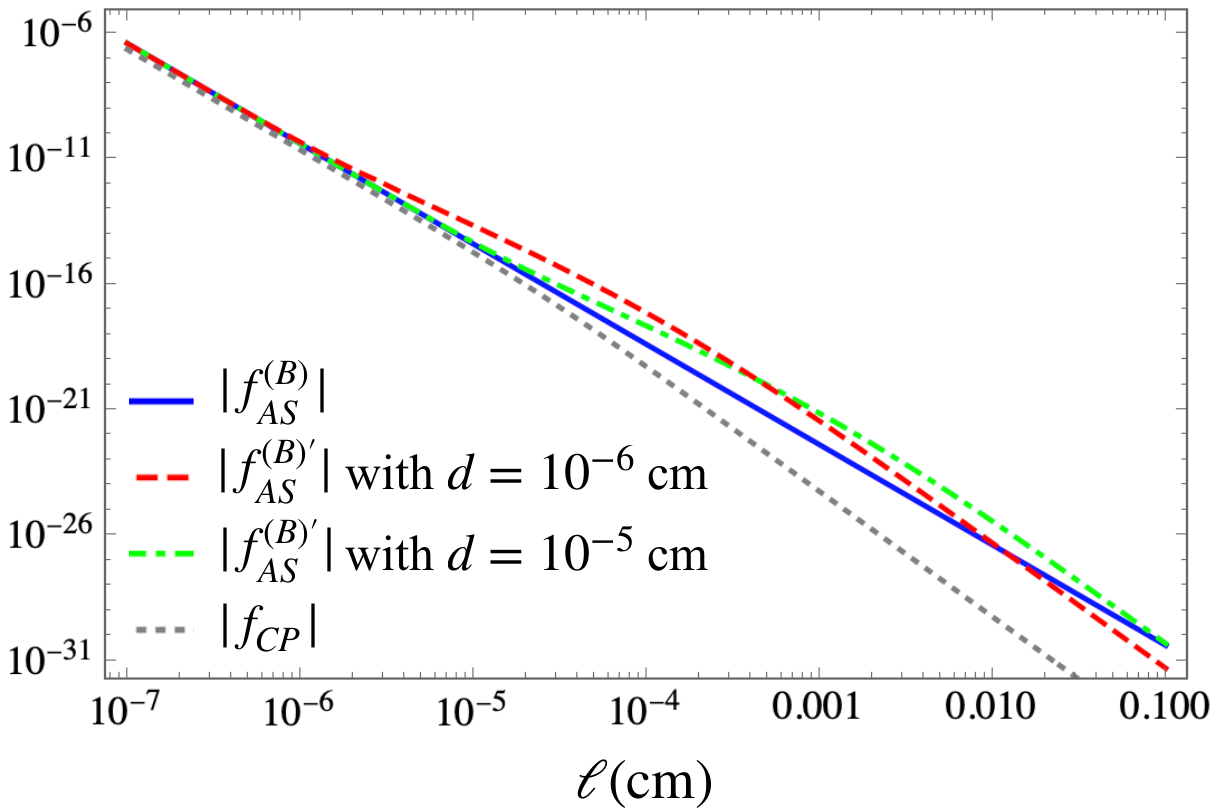}
      \caption{Behavior of the magnitude of the force (dyn) 
      acting on an atom as a function of atom-slab separation $\ell$ (cm):  
            (i)~$|f_{AS}^{(B)}|$ for the system shown in Fig.~\ref{setup-atom-singlelayer}(c) with $\varepsilon_1=1000$ and $\Gamma_{B} = 4.4\times 10^{-8} \, {\rm esu^2/cm}$ (blue curve); 
            (ii)~$|f_{AS}^{(B) \prime}|$ for the system shown in Fig.~\ref{setup-atom-bilayer}(b)  with $\varepsilon_1=1000$, $\varepsilon_{1'} = 11.74$, $\Gamma_{B} = 4.4\times 10^{-8} \, {\rm esu^2/cm}$ and $d = 10^{-6}$ cm (red dashed curve); 
            (iii)~$|f_{AS}^{(B) \prime}|$ for the system shown in Fig.~\ref{setup-atom-bilayer}(b) with $\varepsilon_1=1000$, $\varepsilon_{1'} = 11.74$, $\Gamma_{B} = 4.4\times 10^{-8} \, {\rm esu^2/cm}$ and $d = 10^{-5}$ cm (green dot-dashed curve); 
            (iv)~magnitude of the Casimir-Polder force $|f_{CP}|$ on an atom above a semi-infinite ideal conductor. 
             As the point dipole approximation works well for $\ell$ larger than the typical polar nanoregion size, the small $\ell$ region is an extrapolation.} 
      \label{fAS-plots}
\end{figure} 

Finally, we consider the behavior of the bulk dipolar disorder-induced force acting on an unpolarized, neutral ground-state atom near a slab. 
We have plotted the force behavior in Fig.~\ref{fAS-plots}. 
The behavior for a single-layered and semi-infinite slab system as shown in Fig.~\ref{setup-atom-singlelayer}(c), with $\varepsilon_1 = 1000$ and $\Gamma_{B} = 4.4\times 10^{-8} \, {\rm esu^2/cm}$, is represented by the blue curve. 
The behavior for a bilayered slab system as shown in Fig.~\ref{setup-atom-bilayer}(b), with $\varepsilon_1=1000$, $\varepsilon_{1'} = 11.74$, $\Gamma_{B} = 4.4\times 10^{-8} \, {\rm esu^2/cm}$ and $d = 10^{-6}$ cm,
 is represented by the red dashed curve, whilst the behavior for the same system with $d = 10^{-5}$ cm and all other parameter values unchanged is represented by the green dot-dashed curve. 
For all the cases, we take metastable helium ${\rm{He^\ast}}$ as our atom, whose static polarizability is $\alpha_0 = 315.63 \, {\rm a.u.} = 4.678\times 10^{-23} \, {\rm cm^{3}}$. 
We have calculated the bulk disorder-induced force $f_{AS}^{(B)}$ on an atom near a single-layered slab using Eq.~(\ref{fAS-singlelayer}), whilst the bulk disorder-induced force $f_{AS}^{(B) \prime}$ on an atom near a bilayered slab is calculated using Eq.~(\ref{fAS'}). 

\

We note firstly that for values of $\ell$ smaller than $d$, the disorder-induced  force behavior of $f_{AS}^{(B)\prime}$ tends asymptotically to that of $f_{AS}^{(B)}$ (i.e., the red dashed and green dot-dashed curves converge onto the blue curve), which decays with $\ell^{-4}$. The convergence is expected as the bilayered slab effectively resembles a semi-infinite single-layered slab if the separation distance is sufficiently short. 
Secondly, as we show in Appendix~\ref{app:C}, for separation distances $\ell$ much greater than $d$, the scaling behavior of $f_{AS}^{(B)\prime}$ changes to $\ell^{-5}$. At such distances, the layer containing the quenched bulk dipolar disorder appears very thin, and so the disorder effectively appears to be localized on the surface of its underlying substrate. 
Thirdly, we again note the phenomenon that $f_{AS}^{(B)\prime}$ can become more attractive than $f_{AS}^{(B)}$ if the gap width is sufficiently small, and the range over which $f_{AS}^{(B)\prime}$ is more attractive is larger if $d$ is larger. 
From Fig.~\ref{fAS-plots}, we see that if $d = 10^{-6}$ cm (red dashed curve), $f_{AS}^{(B)\prime}$ is more attractive for $\ell < 0.012$ cm, whilst for the same system with $d = 10^{-5}$ cm (green dot-dashed curve), $f_{AS}^{(B)\prime}$ is more attractive for $\ell < 0.12$ cm. 

\

Lastly, we can compare the behavior of the quenched bulk dipolar disorder-induced force on an unpolarized ground-state atom above a slab with that of the zero-temperature Casimir-Polder force on the same type of atom above an ideal conductor, as the latter provides the theoretical upper bound on the possible magnitude of the Casimir-Polder attraction. 
The zero-temperature Casimir-Polder force on a metastable helium ${\rm{He^\ast}}$ atom positioned at a height $\ell$ above the surface of an ideal conductor is given by 
$
f_{CP} = - \frac{\partial \delta E_{CP}}{\partial \ell}, 
$
where $\delta E_{CP}$ is the Casimir-Polder shift~\cite{wylie1985,lu2025}, viz.,  
\ba
&&\delta E^{CP} 
\!=
-\frac{\hbar c \alpha_0}{4\pi \ell}
\bigg\{
 \bigg(
\frac{\pi}{2\ell^2} \frac{\omega_0}{c} - \frac{\pi \omega_0^3}{c^3}
\bigg) 
\cos\bigg( 2 \frac{\omega_0}{c} \ell \bigg)
\nonumber\\
&&+
\frac{\pi}{\ell} \frac{\omega_0^2}{c^2} 
\sin\bigg( 2 \frac{\omega_0}{c} \ell \bigg)
+ \frac{1}{\ell} \frac{\omega_0^2}{c^2} 
- {\rm Ci}\,
\bigg( 2 \frac{\omega_0}{c} \ell \bigg)
\nonumber\\
&&\times 
\bigg[
\frac{2}{\ell} \frac{\omega_0^2}{c^2} 
\cos\bigg( 2 \frac{\omega_0}{c} \ell \bigg)
-\bigg( 
\frac{1}{\ell^2} \frac{\omega_0}{c} - \frac{\omega_0^3}{c^3}
\bigg)
\sin\bigg( 2 \frac{\omega_0}{c} \ell \bigg)
\bigg]
\nonumber\\
&&- {\rm Si}\,\bigg( 2 \frac{\omega_0}{c} \ell \bigg)
\bigg[
\frac{2}{\ell} \frac{\omega_0^2}{c^2} 
\sin \bigg( 2 \frac{\omega_0}{c} \ell \bigg)
+\bigg( 
\frac{1}{\ell^2} \frac{\omega_0}{c} - 2 \frac{\omega_0^3}{c^3}
\bigg)
\nonumber\\
&&\quad\times 
\cos\bigg( 2 \frac{\omega_0}{c} \ell \bigg)
\bigg]
\bigg\}. 
\nonumber
\ea
Here, $\alpha_0 = 315.63 \, {\rm a.u.} = 4.678\times 10^{-23} \, {\rm cm^{3}}$, $\omega_0 = 1.18 \, {\rm eV} = 1.794\times 10^{15} \, {\rm rad/s}$~\cite{bordag2009}, 
and ${\rm Si}(z) = \int_0^z \frac{\sin t}{t} dt$ and ${\rm Ci}(z) = - \int_z^\infty \frac{\cos t}{t} dt$ are the sine integral and cosine integral respectively. 
The behavior of the Casimir-Polder force is represented by the gray dotted curve in Fig.~\ref{fAS-plots}. We see that for sufficiently large distances, the bulk disorder-induced  force  can be more attractive than the Casimir-Polder force. We also note that both $f_{AS}^{(B) \prime}$ and $f_{CP}$ decay with $\ell^{-4}$ for small values of $\ell$ and they both decay with $\ell^{-5}$ for large values of $\ell$. For $f_{AS}^{(B) \prime}$, the crossover lengthscale is set by $d$, whereas for $f_{CP}$, the crossover lengthscale is set by the dominant transition wavelength of the atom, which is $1.05 \times 10^{-4}$ cm for metastable helium ${\rm{He^\ast}}$~\cite{lu2025}.

\section{Summary and Conclusion} 

In this paper, we have studied the disorder-induced  force between two coplanar layered slabs as well as the disorder-induced  force on an atom above a slab that arise from the presence of quenched random electric dipoles in the slab. 
We have considered the cases for which the quenched dipolar charge disorder resides in the bulk of a given layer as well as on the surface of a given slab, with the electric dipoles being frozen into a disordered glassy state and thermal fluctuations being negligible. 
A key finding is that even in the absence of monopolar charge disorder, if the quenched random polarizations are sufficiently strong, they can give rise to a disorder-induced  force that competes with the Casimir force. As such quenched dipolar charge disorder can be quite significant in materials such as relaxor ferroelectrics, the corresponding disorder-induced  force would have to be taken into account alongside other types of forces (such as the Casimir and Casimir-Polder forces) at the nanoscale. 

\ 

In addition, we have found that for a pair of semi-infinite single-layered slabs separated by a gap of width $\ell$, the disorder-induced pressure decays with $\ell^{-3}$ if the dipolar charge disorder occurs in the bulk, and it decays with $\ell^{-4}$ if the dipolar charge disorder occurs on the slab surface. 
Moreover, we found that the quenched dipolar disorder-induced pressure between the two coplanar slabs (layers ``1" and ``2" separated by a gap layer ``m") can be repulsive if the system obeys the static dielectric contrast rule $\varepsilon_1(0) < \varepsilon_m(0) < \varepsilon_2(0)$ ($\varepsilon_2(0) < \varepsilon_m(0) < \varepsilon_1(0)$), which one may regard as a static analogue of the Dzyaloshinskii-Lifshitz-Pitaevskii (DLP) condition for Casimir-Lifshitz repulsion. This has the following implications: 
(i)~if the system experiences Casimir-Lifshitz repulsion as a result of the Dzyaloshinskii-Lifshitz-Pitaevskii (DLP) condition in the absence of quenched disorder, the said repulsion can be further enhanced if the system contains quenched dipolar disorder that obeys the static dielectric contrast rule, leading to a repulsive quenched dipolar disorder-induced interaction; and 
(ii)~if the static dielectric contrast rule is satisfied in the system with quenched dipolar disorder but the DLP condition is not satisfied, the strength of the Casimir-Lifshitz attraction that the system experiences will be reduced.    

\

If the slabs are separated by a vacuum gap, then the disorder-induced  pressure is always attractive. 
For a system comprising a disorder-free single-layered slab coplanar with and separated by a vacuum gap of width $\ell$ from a disorder-free substrate coated by a film containing bulk dipolar disorder, the dipolar disorder-induced pressure decays with $\ell^{-3}$ ($\ell^{-4}$) for sufficiently small (large) values of $\ell$, the crossover lengthscale between the two asymptotic regimes being set by the thickness of the film that contains the bulk dipolar disorder. 

\ 

For a system consisting of a neutral atom near a single-layered slab containing bulk (surface) dipolar disorder, the disorder-induced force on the atom decays with $\ell^{-4}$ ($\ell^{-5}$) where $\ell$ is the atom-surface separation. If the atom is above a bilayered slab where the top layer contains bulk dipolar disorder, the disorder-induced force decays with $\ell^{-4}$ ($\ell^{-5}$) for sufficiently small (large) values of $\ell$, the crossover lengthscale between the two asymptotia being again set by the thickness of the layer containing the bulk dipolar disorder. 
In all of the above cases considered, the magnitude of the bulk (surface) disorder-induced  force  varies linearly with the mean-square quenched random electric dipole moment per unit volume (area).  

\ 

It is possible to extend the investigation initiated in this paper in a number of directions. For instance, one can consider the effect of higher temperatures and study the effect of partially annealed dipolar disorder on the character of the disorder-induced  force. 
One can also study the effect of dielectric anisotropy and the possibility of a disorder-induced  torque arising from quenched random polarizations. 
Lastly, one may consider a system consisting of coplanar multilayered slabs containing quenched dipolar disorder and separated by a gap, and study how the dipolar disorder-induced  pressure changes as the number of layers and the layer thicknesses are varied.

\section{acknowledgments} 

The author acknowledges financial support from the American University of Sharjah's Faculty Startup Grant (FSU26-S30) and Faculty Research Grant (FRG26-S22). He also thanks David Andelman (Tel Aviv University) for the invitation to contribute to the Special Topic honoring the memory of Rudolf Podgornik, and the JCP editors for their gracious extension of the submission deadline. 
Lastly, the author dedicates this paper to the memory of Rudi, who was not only an inspiring mentor but also an empathetic friend, whose life and actions epitomised the ideal of the humane scientist and Renaissance man.

\begin{widetext}

\appendix

\section{Electrostatic Green functions}
\label{app:A}

Electrostatic Green functions for layered geometries have been obtained for different types of systems~(cf. e.g. Refs.~\cite{schwinger1998,DLP,ninham1970,kitzing1996,netz2000,podgornik2006}. Here, we will derive the Green function that is appropriate to the planar system of Fig.~\ref{setup-slab}. 
Let us write $z_1 = -\ell/2$, $z_m = \ell/2$ and $z_2 = \ell/2+d$ for the $z$-coordinates of the dielectric interfaces. As the Green function $\widetilde{G}(\kv_\parallel, z,z')$ has to vanish as $|z| \to \infty$, we have the following solution Ansatz to Eq.~(\ref{green-eq2}), where we assume that the unit source charge resides at $z = z'$ in layer $2$ (i.e., $z_m < z' < z_2$): 
\ba
\widetilde{G}(\kv_\parallel, z,z')
= \left\{ \begin{array}{ll}
 A_{2'} e^{- k_\parallel z} &
   \quad \mbox{($z > z_2$)},
   \vspace{3mm}\\
A_2 e^{- k_\parallel z} + B_2 e^{k_\parallel z} 
+ \frac{2\pi}{\varepsilon_2 k_\parallel} \, e^{-k|z-z'|} &
   \quad \mbox{($z_m < z < z_2$)},
   \vspace{3mm}\\
A_m e^{- k_\parallel z} + B_m e^{k_\parallel z} &
   \quad \mbox{($z_1 < z < z_m$)},
   \vspace{3mm}\\
B_1 e^{k_\parallel z} &
   \quad \mbox{($z < z_1$)}.  
   \end{array}  \right.
       \label{ansatzg} 
\ea
We have written the Ansatz as a sum of the particular solution $\frac{2\pi}{\varepsilon_2 k} e^{-k|z-z'|}$ and the complementary functions involving $e^{k_\parallel z}$ and $e^{- k_\parallel z}$. The complementary functions allow for the enforcement of the interfacial dielectric boundary conditions, whilst the particular solution accounts for the Dirac delta-function on the right-hand side of Eq.~(\ref{green-eq2}) [as can be verified via the identity $\frac{\partial^2}{\partial z^2} e^{-a|z|} = a^2 e^{-a|z|} - 2 a \delta(z)$]. 
We also have the following boundary conditions: 
\ba
&&\widetilde{G} \Big( \kv_\parallel, z_2+\delta, z' \Big) = \widetilde{G} \Big( \kv_\parallel, z_2-\delta, z' \Big), 
\label{bc1}
\\
&&\left. \varepsilon_{2'} \frac{\partial \widetilde{G}(\kv_\parallel, z, z')}{\partial z} \right\vert_{z=z_2+\delta}  
= \left. \varepsilon_{2} \frac{\partial \widetilde{G}(\kv_\parallel, z, z')}{\partial z}  \right\vert_{z=z_2-\delta},
\label{bc2}
\\
&&\widetilde{G} \Big( \kv_\parallel, z_m+\delta, z' \Big) = \widetilde{G} \Big( \kv_\parallel, z_m-\delta, z' \Big), 
\label{bc3}
\\
&&\left. \varepsilon_2 \frac{\partial \widetilde{G}(\kv_\parallel, z, z')}{\partial z} \right\vert_{z=z_m+\delta}  
= \left. \varepsilon_m \frac{\partial \widetilde{G}(\kv_\parallel, z, z')}{\partial z}  \right\vert_{z=z_m-\delta},
\label{bc4}
\\
&&\widetilde{G} \Big( \kv_\parallel, z_1+\delta, z' \Big) = \widetilde{G} \Big( \kv_\parallel, z_1-\delta, z' \Big), 
\label{bc5}
\\
&&\left. \varepsilon_m \frac{\partial \widetilde{G}(\kv_\parallel, z, z')}{\partial z} \right\vert_{z=z_1+\delta}  
= \left. \varepsilon_1 \frac{\partial \widetilde{G}(\kv_\parallel, z, z')}{\partial z}  \right\vert_{z=z_1-\delta}. 
\label{bc6}
\ea
By solving the boundary condition equations~(\ref{bc5}) and (\ref{bc6}), we obtain
\ba
A_m &\!\!=\!\!& \frac{1}{2} \left( 1-\frac{\varepsilon_1}{\varepsilon_m} \right) e^{2 k_\parallel z_1} B_1, 
\\
B_m &\!\!=\!\!& \frac{1}{2} \left( 1+\frac{\varepsilon_1}{\varepsilon_m} \right) B_1. 
\ea
Using the above two equations in conjunction with the boundary condition equations~(\ref{bc3}) and (\ref{bc4}) leads to 
\ba
A_2 
&\!\!=\!\!& 
\frac{B_1}{4} 
\left( 1+\frac{\varepsilon_m}{\varepsilon_2} \right) \left( 1+\frac{\varepsilon_1}{\varepsilon_m} \right)
\left[
R_{m1} e^{2 k_\parallel z_1} + R_{2m} e^{2 k_\parallel z_m} 
\right], 
\label{eqA2}
\\
B_2 
&\!\!=\!\!& 
\frac{B_1}{4} 
\left( 1+\frac{\varepsilon_m}{\varepsilon_2} \right) \left( 1+\frac{\varepsilon_1}{\varepsilon_m} \right)
\left[
R_{2m} R_{m1} e^{2 k_\parallel (z_1 - z_m)} + 1 
\right]
- \frac{2\pi}{\varepsilon_2 k_\parallel} e^{- k_\parallel z'}.  
\label{eqB2}
\ea
In the above, $R_{{\lambda }{\lambda' }} \equiv \frac{\varepsilon_{\lambda }-\varepsilon_{\lambda' }}{\varepsilon_{\lambda }+\varepsilon_{\lambda' }}$ are the static reflection coefficients and $\varepsilon_{\lambda }$ are the static dielectric permittivities of the ${\lambda }$th layer.
Furthermore, the boundary condition equations~(\ref{bc1}) and (\ref{bc2}) lead to 
\ba
A_2 &\!\!=\!\!& 
\frac{1}{2} \left( 1+\frac{\varepsilon_{2'}}{\varepsilon_2} \right) A_{2'} 
- \frac{2\pi}{\varepsilon_2 k_\parallel} e^{k_\parallel z'}, 
\label{eqA2'}
\\
B_2 &\!\!=\!\!& 
\frac{1}{2} \left( 1-\frac{\varepsilon_{2'}}{\varepsilon_2} \right) e^{- 2 k_\parallel z_2} A_{2'}.  
\label{eqB2'}
\ea
Equating Eq.~(\ref{eqA2}) to Eq.~(\ref{eqA2'}) and Eq.~(\ref{eqB2}) to Eq.~(\ref{eqB2'}) leads to 
\ba
&&\frac{1}{2} \left( 1+\frac{\varepsilon_{2'}}{\varepsilon_2} \right) A_{2'} 
=
\frac{B_1}{4} 
\left( 1+\frac{\varepsilon_m}{\varepsilon_2} \right) \left( 1+\frac{\varepsilon_1}{\varepsilon_m} \right)
\left[
R_{m1} e^{2 k_\parallel z_1} + R_{2m} e^{2 k_\parallel z_m} 
\right] 
+ \frac{2\pi}{\varepsilon_2 k_\parallel} e^{k_\parallel z'}, 
\label{eq1}
\\
&&\frac{1}{2} \left( 1-\frac{\varepsilon_{2'}}{\varepsilon_2} \right) 
e^{- 2 k_\parallel z_2} A_{2'}
=
\frac{B_1}{4} 
\left( 1+\frac{\varepsilon_m}{\varepsilon_2} \right) \left( 1+\frac{\varepsilon_1}{\varepsilon_m} \right)
\left[
R_{2m} R_{m1} e^{2 k_\parallel (z_1 - z_m)} + 1 
\right]
- \frac{2\pi}{\varepsilon_2 k_\parallel} e^{- k_\parallel z'}.
\label{eq2}
\ea
From Eqs.~(\ref{eq1}) and (\ref{eq2}), we obtain  
\be
\frac{B_1}{4} 
\left( 1+\frac{\varepsilon_m}{\varepsilon_2} \right) \left( 1+\frac{\varepsilon_1}{\varepsilon_m} \right)
= 
\frac{
2\pi \big( 
e^{- k_\parallel z'} 
+
R_{22'} \, e^{- k_\parallel (2 z_2 - z')}
\big)  
}
{\varepsilon_2 k_\parallel \big[ 1 - R_{m2} R_{m1} e^{- 2 k_\parallel \ell} 
- R_{22'} \, e^{- 2 k_\parallel d} 
\left(
R_{m1} e^{- 2 k_\parallel \ell} + R_{2m} 
\right) \big]}. 
\ee
Substituting $B_1$ from the above equation into Eqs.~(\ref{eqA2}) and (\ref{eqB2}) allows us to find $A_2$ and $B_2$, viz., 
\ba
A_2 
&\!\!=\!\!& 
\frac{
2\pi \,  e^{2 k_\parallel z_m}  
\big( 
e^{- k_\parallel z'} 
+
R_{22'} \, e^{- k_\parallel (2 z_2 - z')}
\big)  
\left[
R_{2m} + R_{m1} e^{- 2 k_\parallel \ell} 
\right]
}
{\varepsilon_2 k_\parallel \big[ 1 - R_{m2} R_{m1} e^{- 2 k_\parallel \ell} 
- R_{22'} \, e^{- 2 k_\parallel d} 
\left(
R_{m1} e^{- 2 k_\parallel \ell} + R_{2m} 
\right) \big]}, 
\\
B_2 
&\!\!=\!\!& 
\frac{
2\pi \big( 
e^{- k_\parallel z'} 
+
R_{22'} \, e^{- k_\parallel (2 z_2 - z')}
\big)  
\left[
1 - R_{m2} R_{m1} e^{- 2 k_\parallel \ell} 
\right]
}
{\varepsilon_2 k_\parallel \big[ 1 - R_{m2} R_{m1} e^{- 2 k_\parallel \ell} 
- R_{22'} \, e^{- 2 k_\parallel d} 
\left(
R_{m1} e^{- 2 k_\parallel \ell} + R_{2m} 
\right) \big]}
- \frac{2\pi}{\varepsilon_2 k_\parallel} e^{- k_\parallel z'}.  
\ea
Substituting the above values of $A_2$ and $B_2$ into Eq.~(\ref{ansatzg}) leads to Eq.~(\ref{green-multilayer}), which is the Green function for two unit source charges with positions $z$ and $z'$ located within the subspace $\frac{\ell}{2} < z < \frac{\ell}{2} + d$.

\section{Regularization} 
\label{app:reg}

We first show the steps for regularizing 
the integral 
$\int_{L2}\!d^3\rv \, {\bm\nabla}\cdot{\bm\nabla}' G(\rv,\rv') |_{\rv'\to\rv}$ in the {\em bulk} dipolar disorder-averaged electrostatic energy $\overline{U_E} = 
\frac{1}{2} \Gamma_{B} \! \int_{L2}\!d^3\rv \, {\bm\nabla}\cdot{\bm\nabla}' G(\rv,\rv') |_{\rv'\to\rv}$ for the system consisting of a bilayered slab and a single-layered slab separated by a gap of width $\ell$, depicted in Fig.~\ref{setup-slab}(b). 
The relevant Green function is given by Eq.~(\ref{green-multilayer}) in Fourier space. Calling the integral $\mathcal{I}_B(\ell)$, we have 
\ba
\mathcal{I}_B(\ell) &\!\!\equiv\!\!&
\int\!d^3\rv \, {\bm\nabla}\cdot{\bm\nabla}' G(\rv,\rv') |_{\rv'\to\rv}
\nonumber\\
&\!\!=\!\!&
\int\!d^3\rv 
\int \!\! \frac{d^2\kv_\parallel}{(2\pi)^2} 
{\bm\nabla}\cdot{\bm\nabla}' 
e^{i\kv_\parallel\cdot(\rv_\parallel - \rv'_\parallel)} 
\widetilde{G}(\kv_\parallel,z,z') |_{\rv'\to\rv}
\nonumber\\
&\!\!=\!\!&
\mathcal{S} \int_{\ell/2}^{\ell/2+d} \!\!\!\!\!\! dz 
\int \!\! \frac{d^2\kv_\parallel}{(2\pi)^2} 
\left[
\left( k_\parallel^2 + \frac{\partial}{\partial z} \frac{\partial}{\partial z'} \right) 
\widetilde{G}(\kv_\parallel,z,z') 
\right]_{z' \to z}.
\ea
The operator $( k_\parallel^2 + \frac{\partial}{\partial z} \frac{\partial}{\partial z'} )$ eliminates terms involving $e^{-k_\parallel |z-z'|}$, as $\frac{\partial}{\partial z} \frac{\partial}{\partial z'} e^{-k_\parallel |z-z'|} = - k_\parallel^2 e^{-k_\parallel |z-z'|}$. 
At this juncture, let us define the following functions which will enable us to express our formulas more compactly: 
\ba
\Delta &\!\!\equiv\!\!& 1 - R_{1m}R_{2m} e^{-2k_\parallel \ell}, 
\\
\Theta_1 &\!\!\equiv\!\!& R_{1m}  - R_{2m} e^{-2k_\parallel \ell}, 
\label{Theta1-def}
\\
\Theta_2 &\!\!\equiv\!\!& R_{2m}  - R_{1m} e^{-2k_\parallel \ell}. 
\ea
We thus have  
\ba
\mathcal{I}_B(\ell) 
&\!\!=\!\!&
\frac{2\pi \mathcal{S}}{\varepsilon_2} 
\int_{\ell/2}^{\ell/2+d} \!\!\!\!\!\! dz 
\int \!\! \frac{d^2\kv_\parallel}{(2\pi)^2} 
\left[ 
\left( k_\parallel^2 + \frac{\partial}{\partial z} \frac{\partial}{\partial z'} \right) 
\frac{
\Theta_2 \, e^{- k_\parallel (z+z'-\ell)} 
+ \Delta \, R_{22'} \, e^{k_\parallel (z + z' - \ell - 2d)} }
{k_\parallel 
\big(
\Delta
 -
\Theta_2 R_{22'} \, e^{- 2 k_\parallel d} 
\big)} 
\right]_{z' \to z}
\nonumber\\
&\!\!=\!\!&
\frac{\mathcal{S}}{\varepsilon_2}
\int_0^\infty \!\!\! dk_\parallel \, k_\parallel 
\frac{ (1 - e^{-2k_\parallel d}) (\Theta_2 + \Delta R_{22'}) }
{\Delta - \Theta_2 R_{22'} \, e^{- 2 k_\parallel d} }.
\ea
The integral above is divergent and can be regularized by subtracting off the contribution that stems from taking the limit $\ell \to \infty$. As $\ell \to \infty$, $\Delta \to 1$ and $\Theta_2 \to R_{2m}$. Calling the regularized integral $\mathcal{I}^{\rm{reg}}_B$, we have 
\be
\mathcal{I}^{\rm{reg}}_B(\ell) 
\equiv 
\mathcal{I}_B(\ell) - \mathcal{I}_B(\ell\to\infty)
=
\frac{\mathcal{S}}{\varepsilon_2}
\int_0^\infty \!\!\! dk_\parallel \, k_\parallel 
(1 - e^{-2k_\parallel d}) 
\frac{(\Theta_2 - \Delta R_{2m}) (1 + R_{22'}^2 e^{- 2 k_\parallel d})}
{(\Delta - \Theta_2 R_{22'} \, e^{- 2 k_\parallel d}) 
(1 - R_{2m} R_{22'} \, e^{- 2 k_\parallel d})}.  
\label{Ireg}
\ee
This leads to the bulk disorder-averaged electrostatic energy of the quenched random dipoles, Eq.~(\ref{USSB'}):
\ba
U_{SS}^{(B) \prime} 
&\!\!=\!\!& 
\frac{\Gamma_B \mathcal{S}}{2 \varepsilon_2}
\int_0^\infty \!\!\! dk_\parallel \, k_\parallel 
\frac{(1 - e^{-2k_\parallel d}) (\Theta_2 - \Delta R_{2m}) (1 + R_{22'}^2 e^{- 2 k_\parallel d})}
{(\Delta - \Theta_2 R_{22'} \, e^{- 2 k_\parallel d}) 
(1 - R_{2m} R_{22'} \, e^{- 2 k_\parallel d})}.  
\ea
The corresponding regularized integral for a system comprising two single-layered slabs separated by a gap of width $\ell$ (cf. Fig.~\ref{setup-slab}(a)) can be obtained by taking the limit $d\to \infty$ in Eq.~(\ref{Ireg}): 
\be
\mathcal{I}^{\rm{reg}}_B(\ell) 
=
- \frac{\mathcal{S}}{\varepsilon_2}
\int_0^\infty \!\!\! dk_\parallel \, k_\parallel 
\frac{R_{1m} (1-R_{2m}^2) e^{-2k_\parallel \ell}}
{1 - R_{1m} R_{2m} \, e^{- 2 k_\parallel \ell}}
=
- \frac{(1-R_{2m}^2) \, {\rm Li}_2(R_{1m}R_{2m})}{4 \varepsilon_2 R_{2m} \ell^2} \mathcal{S}
\ee
This leads to the bulk disorder-averaged electrostatic energy of the quenched random dipoles, Eq.~(\ref{USSB}):
\be
U_{SS}^{(B)}
= \frac{\Gamma_{B} (1-R_{m2}^2) \, {{\rm Li}}_2(R_{m1} R_{m2})}
{8 \varepsilon_2 R_{m2} \ell^2} \mathcal{S}. 
\ee
Next, we show the details for regularizing 
the integral 
$\int_{L2} \!d^2\rv_\parallel \, {\bm\nabla}\cdot{\bm\nabla}' G(\rv_\parallel,z; \rv'_\parallel,z') |_{\rv'_\parallel\to\rv_\parallel; z,z'\to \ell/2}$ in the {\em surface} dipolar disorder-averaged electrostatic energy 
$\overline{U_E} = \frac{1}{2} \Gamma_{S} \! \int_{L2} \!d^2\rv_\parallel \, {\bm\nabla}\cdot{\bm\nabla}' G(\rv_\parallel,z; \rv'_\parallel,z') |_{\rv'_\parallel\to\rv_\parallel; z,z'\to \ell/2}$
 for the system consisting of a bilayered slab and a single-layered slab shown in Fig.~\ref{setup-slab}(b), 
for the case where the surface disorder is orientationally isotropic. 
The relevant Green function is given by Eq.~(\ref{green-multilayer}) in Fourier space. Calling the integral $\mathcal{I}_S(\ell)$, we have 
\be
\mathcal{I}_S(\ell) 
\equiv
\int_{L2} d^2\rv_\parallel \, {\bm\nabla}\cdot{\bm\nabla}' G(\rv_\parallel,z; \rv'_\parallel,z') |_{\rv'_\parallel\to\rv_\parallel; z,z'\to \ell/2}
=
\mathcal{S} 
\int_{L2} \! \frac{d^2\kv_\parallel}{(2\pi)^2} 
\left( k_\parallel^2 + \frac{\partial}{\partial z} \frac{\partial}{\partial z'} \right) 
\widetilde{G}(\kv_\parallel,z,z') |_{z,z' \to \ell/2}.
\ee
As before, the operator $( k_\parallel^2 + \frac{\partial}{\partial z} \frac{\partial}{\partial z'} )$ eliminates terms involving $e^{-k_\parallel |z-z'|}$. We thus obtain 
\ba
\mathcal{I}_S(\ell)
&\!\!=\!\!&
\frac{2\pi \mathcal{S}}{\varepsilon_2} \!   
\int_{L2} \! \frac{d^2\kv_\parallel}{(2\pi)^2} 
\bigg[
\left( k_\parallel^2 + \frac{\partial}{\partial z} \frac{\partial}{\partial z'} \right) 
\frac{
\Theta_2 \, e^{- k_\parallel (z+z'-\ell)} 
+ \Delta \, R_{22'} \, e^{k_\parallel (z + z' - \ell - 2d)} }
{k_\parallel 
\big(
\Delta
 -
\Theta_2 R_{22'} \, e^{- 2 k_\parallel d} 
\big)} 
\bigg]_{z,z' \to \ell/2}
\nonumber\\
&\!\!=\!\!&
\frac{2 \mathcal{S}}{\varepsilon_2} 
\int_0^\infty \! dk_\parallel \, k_\parallel^2 \, 
\frac{
\Theta_2 + \Delta \, R_{22'} \, e^{- 2 k_\parallel d} }
{\Delta - \Theta_2 R_{22'} \, e^{- 2 k_\parallel d}}. 
\ea
The integral is divergent. We can regularize it by subtracting the contribution coming from the limit $\ell\to\infty$:
\be
\mathcal{I}^{{\rm reg}}_S(\ell)
\equiv 
\mathcal{I}_S(\ell) - \mathcal{I}_S(\ell\to\infty) 
=
\frac{2 \mathcal{S}}{\varepsilon_2} 
\int_0^\infty \! dk_\parallel \, k_\parallel^2 
\frac{(\Theta_2 - \Delta R_{2m}) (1 + R_{22'}^2 \, e^{- 4 k_\parallel d})}
{(\Delta - \Theta_2 R_{22'} \, e^{- 2 k_\parallel d}) (1 - R_{2m}R_{22'} \, e^{- 2 k_\parallel d})}. 
\label{reg1}
\ee
This leads to the surface dipolar disorder-averaged electrostatic energy for a system comprising a single-layered slab and a bilayered slab: 
\be
U_{SS}^{(S) \prime} = 
\frac{\Gamma_S \mathcal{S}}{\varepsilon_2} 
\int_0^\infty \! dk_\parallel \, k_\parallel^2 
\frac{(\Theta_2 - \Delta R_{2m}) (1 + R_{22'}^2 \, e^{- 4 k_\parallel d})}
{(\Delta - \Theta_2 R_{22'} \, e^{- 2 k_\parallel d}) (1 - R_{2m}R_{22'} \, e^{- 2 k_\parallel d})}. 
\label{Usss1}
\ee
For the case where the surface disorder is orientationally anisotropic, the disorder-averaged electrostatic energy is given by Eq.~(\ref{energy-S-aniso}), and we need to regularize the integrals $\int_{L2} \!d^2\rv_\parallel \, {\bm\nabla}_\parallel \!\cdot\! {\bm\nabla}'_\parallel 
G(\rv_\parallel,z; \rv'_\parallel,z') |_{\rv'_\parallel\to\rv_\parallel; z,z'\to \ell/2}$ 
and 
$\int_{L2} \!d^2\rv_\parallel \, \frac{\partial}{\partial z} \frac{\partial}{\partial z'}  
G(\rv_\parallel,z; \rv'_\parallel,z') |_{\rv'_\parallel\to\rv_\parallel; z,z'\to \ell/2}$. 
By doing a calculation analogous to the previous one, it can be shown that the regularized expression for $\int_{L2} \!d^2\rv_\parallel \, {\bm\nabla}_\parallel \!\cdot\! {\bm\nabla}'_\parallel 
G(\rv_\parallel,z; \rv'_\parallel,z') |_{\rv'_\parallel\to\rv_\parallel; z,z'\to \ell/2}$ 
is the same as the regularized expression for
$\int_{L2} \!d^2\rv_\parallel \, \frac{\partial}{\partial z} \frac{\partial}{\partial z'}  
G(\rv_\parallel,z; \rv'_\parallel,z') |_{\rv'_\parallel\to\rv_\parallel; z,z'\to \ell/2}$, given by $\frac{1}{2}\mathcal{I}^{{\rm reg}}_S(\ell)$. Thus we obtain for the surface dipolar disorder-averaged electrostatic energy the following expression:  
\be
U_{SS}^{(S) \prime} = 
\frac{(\Gamma_S^\parallel + \Gamma_S^\perp) \mathcal{S}}{2\varepsilon_2} 
\int_0^\infty \! dk_\parallel \, k_\parallel^2 
\frac{(\Theta_2 - \Delta R_{2m}) (1 + R_{22'}^2 \, e^{- 4 k_\parallel d})}
{(\Delta - \Theta_2 R_{22'} \, e^{- 2 k_\parallel d}) (1 - R_{2m}R_{22'} \, e^{- 2 k_\parallel d})}. 
\label{Usss2}
\ee
Equations~(\ref{Usss1}) and (\ref{Usss2}) can be expressed by the following equation 
\be
U_{SS}^{(S) \prime} = 
\frac{\Gamma_S^\ast \mathcal{S}}{\varepsilon_2} 
\int_0^\infty \! dk_\parallel \, k_\parallel^2 
\frac{(\Theta_2 - \Delta R_{2m}) (1 + R_{22'}^2 \, e^{- 4 k_\parallel d})}
{(\Delta - \Theta_2 R_{22'} \, e^{- 2 k_\parallel d}) (1 - R_{2m}R_{22'} \, e^{- 2 k_\parallel d})}, 
\ee
where $\Gamma_S^\ast = \Gamma_S$ if the surface disorder is orientationally isotropic with disorder statistics given by Eq.~(\ref{stats-S}), and $\Gamma_S^\ast = \frac{1}{2}(\Gamma_S^\parallel + \Gamma_S^\perp)$ if the surface disorder is orientationally anisotropic with disorder statistics given by Eq.~(\ref{stats-S-aniso}). 

\

We can take the limit $d\to\infty$ in the above result to obtain the surface dipolar disorder-averaged electrostatic energy for a system comprising two single-layered slabs separated by a gap medium:
\be
U_{SS}^{(S) \prime}
=
- \frac{\Gamma_S^\ast \mathcal{S}}{\varepsilon_2} 
\int_0^\infty \! dk_\parallel \, k_\parallel^2 \, 
\frac{R_{1m} (1 - R_{2m}^2) \, e^{-2k_\parallel \ell}}
{1 - R_{1m} R_{2m} e^{-2k_\parallel \ell}}
=
\frac{\Gamma_S^\ast(1-R_{m2}^2) \, {\rm Li}_3(R_{m1}R_{m2})}
{4 \varepsilon_2 R_{m2} \ell^3} \mathcal{S}, 
\ee
which is Eq.~(\ref{USSS}). In the above, we have used the identity
\be
\int_0^\infty \!\!\! dk_\parallel \, k_\parallel^2 
\left( 
\frac{e^{- 2 k_\parallel \ell}}
{A-B e^{-2k_\parallel \ell}}
\right) 
= \frac{{\rm Li}_3 \big(\frac{B}{A}\big)}{4B\ell^3}. 
\label{identity}
\ee

\section{Asymptotic expressions for $d \ll \ell$} 
\label{app:C}

We first consider Eq.~(\ref{USSB'}), which is the regularized disorder-averaged electrostatic energy of quenched random electric dipoles for the system consisting of a bilayered slab and a single-layered slab depicted in Fig.~\ref{setup-slab}(b). For $d \ll \ell$, the energy can be approximated by 
\ba
U_{SS}^{(B) \prime}
&\!\!\approx\!\!&  
\frac{\Gamma_B d \mathcal{S} (1 + R_{22'}^2)}{\varepsilon_2 (1 - R_{2m} R_{22'})}
\int_0^\infty \!\!\! dk_\parallel \, k_\parallel^2 \,  
\frac{\Theta_2 - \Delta R_{2m}}
{\Delta - \Theta_2 R_{22'}}.
\ea
Using Eq.~(\ref{identity}), we find 
\be
U_{SS}^{(B) \prime}
\approx 
- \frac{\Gamma_B d (1 + R_{22'}^2) (1-R_{2m}^2) \, {\rm Li}_3 \Big[ \frac{R_{1m} (R_{2m} - R_{22'})}{1-R_{22'} R_{2m}} \Big] }
{4 \varepsilon_2 (1 - R_{2m} R_{22'}) (R_{2m} - R_{22'}) \ell^3} \mathcal{S}. 
\ee
The corresponding disorder-induced pressure for $d \ll \ell$ is 
\be
\mathcal{P}_{SS}^{(B) \prime}
\approx 
- \frac{3 \Gamma_B d (1 + R_{22'}^2) (1-R_{2m}^2) \, {\rm Li}_3 \Big[ \frac{R_{1m} (R_{2m} - R_{22'})}{1-R_{22'} R_{2m}} \Big] }
{4 \varepsilon_2 (1 - R_{2m} R_{22'}) (R_{2m} - R_{22'}) \ell^4}, 
\ee
which decays as $\ell^{-4}$.  
If we set $\varepsilon_{2'} = \varepsilon_2$, we recover Eq.~(\ref{USSS}), provided that we identify $\Gamma_S^\ast = \Gamma_{B} d$. 

\

Next, we consider the disorder-induced  force  acting on an atom in the vacuum positioned a perpendicular distance $\ell$ above the surface of a bilayered slab which contains bulk dipolar disorder in the layer adjacent to the vacuum subspace, Eq.~(\ref{fAS'}). For $d \ll \ell$, the force can be approximated by 
\be
f_{AS}^{(B) \prime} 
=
- \frac{32 \pi \alpha_0 (1 + R_{11'}^2) \, \Gamma_B d}
{(\varepsilon_1+1)^2 (1 - R_{1m}R_{11'})^2}
\int_0^\infty \!\!\!\!\! dk_\parallel \, 
k_\parallel^4 \, e^{-2k_\parallel\ell} 
=
- \frac{24 \pi \alpha_0 (1 + R_{11'}^2) \, \Gamma_B d}
{(\varepsilon_1+1)^2 (1 - R_{1m}R_{11'})^2 \ell^5}. 
\label{fASthin1}
\ee
The disorder-induced  force thus decays with $\ell^{-5}$, which is the power-law decay behavior expected for the case of quenched dipolar disorder localized on the surface of layer 1. 
As a check of consistency, we can set $\varepsilon_{1'} = \varepsilon_1$. The above becomes 
\be
f_{AS}^{(B) \prime} 
= 
- \frac{24 \pi \alpha_0 \Gamma_B d}{(\varepsilon_1+1)^2 \ell^5}, 
\label{fASthin2}
\ee
which is Eq.~(\ref{fASS}) if we identify $\Gamma_B d = \Gamma_S^\ast$. 

\end{widetext}

\end{document}